\documentclass[aps,pre,reprint,superscriptaddress]{revtex4-2}

\usepackage[a4paper, top=1.6cm, bottom=1.6cm, left=1.4cm, right=1.4cm, columnsep=0.6cm]{geometry}

\usepackage[utf8]{inputenc}
\usepackage[T1]{fontenc}

\usepackage{amsmath, amssymb, amsthm}
\usepackage{graphicx}
\usepackage{bm}
\usepackage{color}
\usepackage{cancel}
\usepackage{tikz}
\usetikzlibrary{graphs}
\usepackage{caption}
\usepackage{microtype}
\usepackage{lmodern}
\usepackage{braket}
\usepackage{blkarray}

\DeclareMathOperator{\Roff}{R_{off}}
\DeclareMathOperator{\Ron}{R_{on}}

\newcommand{\comm}[1]{}

\newtheorem*{theorem*}{Theorem}

\begin{document}

\title{\textbf{A Unifying Approach to Self-Organizing Systems Interacting via Conservation Laws}}

\author{F. Barrows}
\email{fbarrows@lanl.gov}
\affiliation{Theoretical Division (T4), Los Alamos National Laboratory, Los Alamos, New Mexico 87545, USA}
\affiliation{Center for Nonlinear Studies, Los Alamos National Laboratory, Los Alamos, New Mexico 87545, USA}

\author{G. Zhang}
\email{gz2241@nyu.edu}
\affiliation{Center for Soft Matter Physics, New York University, 726 Broadway, New York, NY 10003, USA}
\affiliation{Simons Center for Computational Physical Chemistry, Department of Chemistry, New York University, New York, NY 10003, USA}

\author{S. Anand}
\affiliation{Courant Institute, New York University, 251 Mercer St, New York, NY 10012, USA}

\author{Z. Chen}
\affiliation{Courant Institute, New York University, 251 Mercer St, New York, NY 10012, USA}

\author{J. Lin}
\affiliation{Theoretical Division (T4), Los Alamos National Laboratory, Los Alamos, New Mexico 87545, USA}
\affiliation{Center for Nonlinear Studies, Los Alamos National Laboratory, Los Alamos, New Mexico 87545, USA}

\author{A. Desai}
\affiliation{Center for Nonlinear Studies, Los Alamos National Laboratory, Los Alamos, New Mexico 87545, USA}
\affiliation{University of California, Berkeley, Department of Electrical Engineering and Computer Sciences, Berkeley, CA 94720, USA}
\affiliation{Computer, Computational, and Statistical Sciences (CCS-3), Los Alamos National Laboratory, Los Alamos, New Mexico 87545, USA}

\author{S. Martiniani}
\email{stefano.martiniani@nyu.edu}
\affiliation{Center for Soft Matter Physics, New York University, 726 Broadway, New York, NY 10003, USA}
\affiliation{Simons Center for Computational Physical Chemistry, Department of Chemistry, New York University, New York, NY 10003, USA}
\affiliation{Courant Institute, New York University, 251 Mercer St, New York, NY 10012, USA}
\affiliation{Center for Neural Science, New York University, New York, NY 10003, USA}

\author{F. Caravelli}
\email{caravelli@lanl.gov}
\affiliation{Theoretical Division (T4), Los Alamos National Laboratory, Los Alamos, New Mexico 87545, USA}
\affiliation{Center for Nonlinear Studies, Los Alamos National Laboratory, Los Alamos, New Mexico 87545, USA}
\affiliation{Dipartimento di Fisica dell’Università di Pisa, Largo Bruno Pontecorvo 3, I-56127 Pisa, Italy}

\begin{abstract}
We present a unified framework for embedding and analyzing dynamical systems using generalized projection operators rooted in local conservation laws. By representing physical, biological, and engineered systems as graphs with incidence and cycle matrices, we derive dual projection operators that decompose network fluxes and potentials. This formalism aligns with principles of non-equilibrium thermodynamics and captures a broad class of systems governed by flux-forcing relationships and local constraints.

We extend this approach to collective dynamics through the \textit{PRojective Embedding of Dynamical Systems} (PrEDS), which lifts low-dimensional dynamics into a high-dimensional space, enabling both replication and recovery of the original dynamics. When systems fall within the PrEDS class, their collective behavior can be effectively approximated through projection onto a mean-field space.

We demonstrate the versatility of PrEDS across diverse domains, including resistive and memristive circuits, adaptive flow networks (e.g., slime molds), elastic string networks, and particle swarms. Notably, we establish a direct correspondence between PrEDS and swarm dynamics, revealing new insights into optimization and self-organization.

Our results offer a general theoretical foundation for analyzing complex networked systems and for designing systems that self-organize through local interactions.
\end{abstract}

\maketitle

\tableofcontents


\section{Introduction}

Many complex dynamical systems, -from physical circuits and elastic media to biological networks like slime molds and swarms, -exhibit emergent low-dimensional behavior despite operating in high-dimensional spaces. Understanding how such dimensionality reduction arises, especially from first principles, remains a central challenge in nonlinear science and dynamical systems theory. Recent evidence from neuroscience suggests that neural dynamics during resting states lie on low-dimensional manifolds \cite{Cueva_PNAS_2020,Song_elife_2023}, though the mechanisms behind this remain under investigation \cite{Runfola_npj_2025}.

In this work, we introduce a unifying formalism for a broad class of such systems, based on the observation that local conservation laws naturally give rise to projection operators, which constrain dynamics onto subspaces. Our central tool is the \emph{Projective Embedding of Dynamical Systems} (PrEDS) framework \cite{CARAVELLI2023133747}, which formalizes how these projection operators, -derived from graph representations of networks, -define the structure of low-dimensional manifolds and influence emergent collective behavior.

The first goal of this paper is to demonstrate that these projector structures are \textit{ubiquitous} in systems governed by local conservation laws, including electrical circuits, mechanical spring networks, and biological transport systems. The second goal is to show that when a connection to the PrEDS formalism is established, then there is a standard way of deriving mean field theories for these systems. The third goal is to show that, when cast in the PrEDS formalism, these systems exhibit dynamics equivalent to interacting particle swarms, where collective behavior arises through a combination of gradient-following and inter-agent coupling.

In this regard, we show that PrEDS dynamics correspond to a collective optimization process: particles descend along an averaged gradient, while an orthogonal term enforces cohesion via a harmonic attraction. This formulation naturally recovers swarm-like behavior and leads to a continuum description for the evolving particle density, providing a framework to study a large class of complex adaptive systems \cite{CAS}. Through simulations, we show that PrEDS-based swarms more reliably converge to global optima than independent gradient-descent agents.

At the heart of PrEDS, which is an alternative interpretation, is that it can be considered as a lifting of low-dimensional dynamics into a high-dimensional space, coupled with a carefully constructed projection operator (typically derived from the incidence or cycle matrix of a network). This operator cleanly separates reversible and irreversible components of the dynamics and preserves fixed points under projection. The approach thus enables both a microscopic and macroscopic understanding of constrained collective evolution.

Projection-based methods are widely used in dynamical systems, notably in the Mori–Zwanzig (MZ) formalism \cite{Grabert1982,zwanzig2001nonequilibrium} and in projected dynamical systems \cite{Nagurney1996,NAGURNEY2006646,Pang2008}, but PrEDS differs in two essential ways: (1) it emphasizes \emph{intrinsic} local conservation laws rather than external constraints or unresolved variables, and (2) it naturally lends itself to collective mean-field behavior, applicable to both physical and biological systems.

Graph representations provide a powerful lens for unifying these ideas. Many physical networks, such as electrical or elastic systems, can be formulated in terms of graphs where local interactions (e.g., Kirchhoff’s laws) imply specific projector forms. Network science has studied such systems extensively \cite{Barabasi_Science_1999,Barabasi_RevModPhys_2002,BOCCALETTI2006175}, yet detailed connections between topology and dynamics, -particularly in systems with local flow conservation, -remain underdeveloped \cite{PhysRevResearch.2.023352,PhysRevLett.85.4633,Dorogovtsev01062002,en11061381,Hofmann_EuroJClinicalInves_2018}. This work aims to bridge that gap.

To illustrate, consider Kirchhoff’s Current Law (KCL), which enforces local current conservation at nodes. For a directed graph with incidence matrix $B$, KCL requires $B\vec{i} = 0$, where $\vec{i}$ is the edge current vector. This can be reformulated as a projection condition $\boldsymbol{\Omega}_B \vec{i} = 0$, where $\boldsymbol{\Omega}_B = B^t(B B^t)^{-1} B$ is a projector satisfying $\boldsymbol{\Omega}_B^2 = \boldsymbol{\Omega}_B$ \cite{Caravelli_PRE_2017}. This structure emerges generically in time-varying systems subject to conservation laws.

Similar projector constructions arise in mechanical systems (e.g., force balance in spring networks), in fluid systems governed by Poiseuille flow, and in adaptive systems such as memristive circuits. These examples are not coincidental: they all reflect local conservation of flow-like quantities and admit formulations based on dual projection operators derived from graph topology. Throughout this paper, we show how these systems fit within the PrEDS framework and how their dynamics reduce to collective optimization.

The appearance of projectors can also be grounded in nonequilibrium thermodynamics. Classic work by Onsager, Prigogine, Oster, Perelson, and others extended thermodynamic reasoning to networked and driven systems \cite{Onsager_PRA_2931,Prigogine_JPhysChem_1967,OSTER_Nat_1971,Perelson1975-zz,Agren_JPhaseEqui_2022,PhysRevE.90.062131,PhysRevE.73.036126,Jarzynski_PRL_1997}, where entropy production, flux–force relationships, and reciprocal symmetries play key roles. Our approach builds on this foundation, identifying projection operators from local constraints, and using them to analyze how global behavior arises in systems driven by external forcing \cite{chua1976graph,Cederbaum_CircuitGraph_1984,Caravelli_PRE_2017}.

\textit{Structure of the paper:} In Section II, we derive projection operators from network structure and define the PrEDS formalism. In Section III, we apply PrEDS to physical and biological systems, including circuits, flow networks, and spring networks, and compare network-based and mean-field embeddings. In Section IV, we show that PrEDS corresponds to the dynamics of particle swarms and derive a continuum theory for swarm density evolution. Full derivations and extended models are included in the appendices.

\section{Network Dynamics, Projection Operators, and PrEDS}

Reducing high-dimensional dynamics to effective low-dimensional representations remains a central challenge across physical, biological, and engineered systems. This issue appears in many contexts, from neuroscience \cite{Cueva_PNAS_2020,Song_elife_2023,Runfola_npj_2025} to control theory, and from statistical physics to adaptive materials. While many successful data-driven approaches exist for identifying low-dimensional embeddings, these techniques often obscure the physical origin of the reduction itself.

In this work, we argue that low-dimensional dynamics frequently arise from local conservation laws that constrain the evolution of networked systems. These conservation laws, -of charge, mass, momentum, or energy, -define subspaces within which the dynamics are forced to remain. Our goal is to show that such constraints naturally give rise to algebraic projection operators, and that these projectors serve as the structural foundation of a general framework we develop here: the Projective Embedding of Dynamical Systems (PrEDS). The novelty of this approach lies in its explicit construction of projectors from the system’s topology, and in its use of these operators to embed and simulate constrained dynamics in a higher-dimensional space. In this section, we explain how this framework emerges, which build on existing work, and what results we derive.

Projected dynamical systems also play a role in constrained optimization and control theory \cite{Nagurney1996,NAGURNEY2006646,Pang2008}, where projections enforce evolution within convex feasible sets. Both approaches rely on projection as a computational or modeling tool, often decoupled from the system’s physical substrate.

By contrast, the projection operators we consider in this work arise directly from structural properties of the system. They are not added to enforce constraints but emerge from the conservation laws themselves. This distinction places our work closer to the tradition of nonequilibrium thermodynamics \cite{Onsager_PRA_2931,Prigogine_JPhysChem_1967,OSTER_Nat_1971,Perelson1975-zz,Agren_JPhaseEqui_2022,PhysRevE.90.062131,PhysRevE.73.036126}, in which physical constraints and symmetries govern the form of the equations. Our derivations build most directly on the work of Oster and Perelson, who showed how the topology of a network defines the dissipation structure and response functions of chemical and electrical systems \cite{OSTER_Nat_1971,Perelson1975-zz}.

\subsection{Projection Operators from Graph Structure}

To formalize the emergence of projection operators in networked dynamical systems, we begin by considering systems represented as graphs $\mathcal{G} = (V, E)$, where $V$ is a set of $n$ nodes and $E$ is a set of $m$ edges. Nodes represent conservation sites (e.g., junctions), while edges represent dynamic interactions (e.g., currents, flows, or forces). These graphs may describe physical, biological, or engineered systems, and provide a natural structure for expressing conservation laws.

A physical system defined on a graph may be analyzed from either a nodal or a loop perspective. The nodal view centers on how nodes are interconnected, while the loop view examines the cycles formed by edge sequences. These complementary approaches are central in circuit theory, but their applicability is much broader.

To encode node-level conservation, such as Kirchhoff’s Current Law (KCL), we define the incidence matrix $B \in \mathbb{R}^{n \times m}$. Each row of $B$ corresponds to a node, and each column to an edge. The entries $b_{k,e}$ of $B$ are defined as:
\begin{equation}
b_{k,e} = 
\begin{cases}
1 & \text{if node } k \text{ is the source of edge } e, \\
-1 & \text{if node } k \text{ is the target of edge } e, \\
0 & \text{otherwise}.
\end{cases}
\end{equation}
Given a vector of edge flows $\vec{i} \in \mathbb{R}^m$, the node conservation law becomes
\begin{equation}
B \vec{i} = 0,
\end{equation}
ensuring that the net inflow at each node vanishes. Of course, the incidence matrix $B$ can be defined also for other types of systems with conservation laws described by a graph.

Instead, to capture conservation over loops, such as Kirchhoff’s Voltage Law (KVL), we define the cycle matrix $A \in \mathbb{R}^{c \times m}$, where each row represents a cycle in the graph. The entries of $A$ take values $\pm 1$ if an edge is included in the cycle (depending on orientation), and 0 otherwise. Voltage configurations $\vec{v}$ must satisfy:
\begin{equation}
A \vec{v} = 0,
\label{eqn:AvKernel}
\end{equation}
which enforces that the total potential drop around each cycle is zero. Since cycles are not independent, one typically reduces $A$ to a basis of fundamental loops. The construction of both $B$ and $A$ is independent of physical parameters and derives purely from the graph's topology.

These constraints define subspaces of admissible flows and potentials. The space of flows that satisfy node conservation lies in the nullspace of $B$, while the voltages that obey KVL lie in the nullspace of $A$. To project arbitrary vectors into these constraint subspaces, we define the orthogonal projection operators:
\begin{equation}
\boldsymbol{\Omega}_B = B^T (B B^T)^{-1} B, \qquad \boldsymbol{\Omega}_A = A^T (A A^T)^{-1} A.
\end{equation}
Here, $\boldsymbol{\Omega}_B$ projects edge variables (such as voltage or force) onto the subspace compatible with node conservation, while $\boldsymbol{\Omega}_A$ enforces loop conservation constraints. These operators are orthogonal projectors satisfying the relations $\boldsymbol{\Omega}_B^2 = \boldsymbol{\Omega}_B$, $\boldsymbol{\Omega}_A^2 = \boldsymbol{\Omega}_A$, $\boldsymbol{\Omega}_A \boldsymbol{\Omega}_B = 0$, and $\boldsymbol{\Omega}_A + \boldsymbol{\Omega}_B = I$.

In Appendix~\ref{appendix:projoperators}, we provide a detailed derivation of these operators using a simple triangle graph as an illustrative example. There we show how $B$ and $A$ encode fundamental walks and loops, and how Tellegen’s theorem implies the orthogonality of the two constraint subspaces. We also demonstrate how these structures arise naturally from network thermodynamics, where they serve as building blocks for dissipation functionals and kinetic matrices.

Although the projection operators originate from combinatorial graph theory, they acquire physical significance when flows, potentials, or other state variables are introduced. In what follows, we reinterpret these projectors as generators of constrained dynamics and use them as the foundation for the PrEDS framework across multiple physical domains.

In Appendix~\ref{appendix:nettherm}, we provide a complete derivation of these projectors from a network thermodynamics perspective. This includes: (i) a derivation of $\boldsymbol{\Omega}_A$ and $\boldsymbol{\Omega}_B$ from constrained power minimization, (ii) a formulation of entropy production under projector action, and (iii) the appearance of these projectors as kinetic tensors in Onsager’s linear irreversible framework. This formalism recovers familiar results for resistive circuits, but also generalizes to arbitrary flow-conserving systems.

The projectors $\boldsymbol{\Omega}_A$ and $\boldsymbol{\Omega}_B$ form the backbone of the PrEDS framework. While they are mathematically defined by the graph structure, we derive them from physical first principles to establish their universal applicability. Importantly, our derivation does not assume that the system is near equilibrium or time-independent. It applies to driven, time-varying systems with memory, -a key departure from classical linear treatments.

Furthermore, we reinterpret these projectors not just as constraints but as generators of dynamical structure. That is, the presence of a projector defines the subspace in which both evolution and memory update occur. This insight leads directly to the definition of PrEDS.

\subsection{Projective embeddings of dynamical systems}
Let $\vec{x} \in \mathbb{R}^m$ denote the state of a dynamical system with $m$ components, evolving as
\begin{equation}
\frac{d\vec{x}}{dt} = \vec{f}(\vec{x}),
\end{equation}
where $\vec{f} : \mathbb{R}^m \rightarrow \mathbb{R}^m$ is a general (possibly nonlinear) vector field. In the \textit{Projective Embedding of Dynamical Systems} (PrEDS) framework, each scalar variable $x_i$ is lifted to an $N$-dimensional vector $\vec{X}_i \in \mathbb{R}^N$, whose entries we denote by $x_{i\beta}$ with $\beta = 1, \dots, N$. These lifted variables represent replicated or distributed versions of the original degrees of freedom.

To facilitate nonlinear operations in the lifted space, we define the diagonal matrix $\mathbf{X}_i \in \mathbb{R}^{N \times N}$ by setting its diagonal to be $\vec{X}_i$, i.e.,
\begin{equation}
\mathbf{X}_i = \text{Diag}(x_{i1}, x_{i2}, \dots, x_{iN}).
\end{equation}

The dynamics of each lifted variable $\vec{X}_i$ are governed by the equation
\begin{equation}
\frac{d\vec{X}_i}{dt} = \boldsymbol{\Omega} \vec{F}_i(\boldsymbol{\Omega}, \{ \vec{X}_j \}) - \alpha (\mathbf{I} - \boldsymbol{\Omega}) \vec{X}_i,
\label{eq:PrEDS_dynamics}
\end{equation}
where $\boldsymbol{\Omega} \in \mathbb{R}^{N \times N}$ is a projection operator satisfying $\boldsymbol{\Omega}^2 = \boldsymbol{\Omega}$, and $\mathbf{I}$ is the identity matrix of compatible size. The function $\vec{F}_i$ is the lifted counterpart of $f_i$ and acts on the set of lifted variables $\{\vec{X}_j\}_{j=1}^m$. In practice, this function is often evaluated row-wise, where each row corresponds to a replica (or agent) and receives as input a common projected configuration $\boldsymbol{\Omega} \vec{X}_j$ for each $j$.

\begin{figure}[ht]
    \centering
    \includegraphics[width=.6\linewidth]{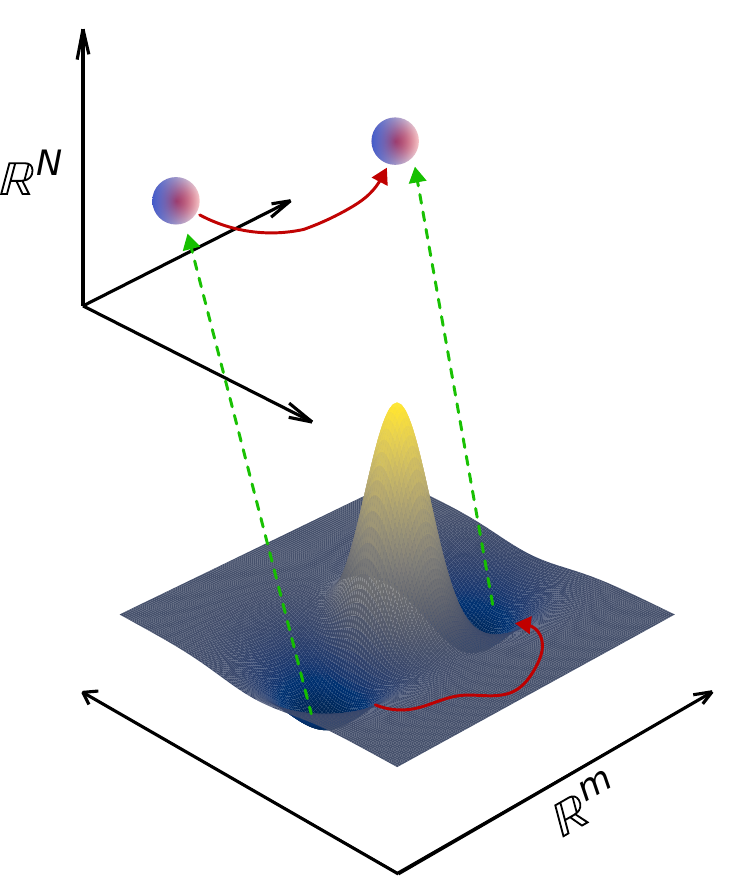}
    \caption{A graphical representation of the PrEDS mapping. PrEDS map a dynamical system $\mathcal O_1$ in $\mathbb R^m$ to a dynamical system $\mathcal O_2$ in $\mathbb R^N$, with the stable fixed points in $\mathcal O_1$ that can be mapped to the stable fixed points in $\mathcal O_2$. This can be seen as the fact that for certain dynamical systems in $\mathbb R^N$ there exists a lower-dimensional dynamical system with an equal number of stable fixed points that can be mapped to the stable fixed points of the original system.}
    \label{fig:PrEDSmap}
\end{figure}

A picture of such mapping is shown in Fig. \ref{fig:PrEDSmap}.

The term $-\alpha (\mathbf{I} - \boldsymbol{\Omega}) \vec{X}_i$ enforces decay of any component orthogonal to the range of $\boldsymbol{\Omega}$, thereby confining the long-time behavior of the system to the subspace defined by the projector. Depending on the context, the projection operator $\boldsymbol{\Omega}$ may correspond to a graph-based projector such as $\boldsymbol{\Omega}_A = A^T (A A^T)^{-1} A$ (based on fundamental cycles) or $\boldsymbol{\Omega}_B = B^T (B B^T)^{-1} B$ (based on node conservation), or to a mean-field projector such as $\boldsymbol{\Omega}_{\text{MF}}$ with entries $\Omega_{\alpha\beta} = 1/N$.

In cases where $\vec{f}$ derives from a potential, i.e., $\vec{f} = -\nabla V$, the function $\vec{F}_i$ typically does not retain this gradient structure in the lifted space. Nevertheless, under the commutative map condition defined in \cite{CARAVELLI2023133747}, the projection acts on the output of the function, not its argument, ensuring that the fixed points of the original system are preserved. This property is formalized in the Banality Lemma \cite{CARAVELLI2023133747}. The original scalar dynamics can be recovered by taking a mean over the lifted variables:
\begin{equation}
x_i(t) = \frac{1}{N} \vec{1}^T \boldsymbol{\Omega} \vec{X}_i(t),
\end{equation}
where $\vec{1}$ is the all-ones vector in $\mathbb{R}^N$. This averaging operation collapses the replicated dynamics back to the original low-dimensional space, preserving asymptotic structure while enabling collective and swarm-like generalizations.

This formulation guarantees that the long-time dynamics remain confined to the projector-defined manifold. Moreover, it preserves the fixed points of the original system, as shown by the Banality Lemma in \cite{CARAVELLI2023133747}. We rigorously derive the embedding, prove fixed-point preservation, and analyze its behavior in linear and nonlinear examples in Appendix~\ref{appendix:PrEDSformulation}.

A particularly important case occurs when $\Omega$ is chosen as the mean-field projector, where every entry $\boldsymbol{\Omega}_{\alpha\beta} = 1/N$. In this case, the system behaves as a swarm: each variable responds to the average state of all others, while being softly attracted toward that average. This type of interaction is ubiquitous in swarm intelligence \cite{Bonabeau1999, Dorigo1997}, collective decision-making, and distributed optimization.

We show below in section~\ref{appendix:meanfieldexamples}, with illustrative examples that PrEDS with mean-field projection leads to collective dynamics capable of escaping local minima and converging to global optima. This is illustrated using multi-well potentials, where swarm-like relaxation under PrEDS outperforms standard gradient descent.

As a result, we aim to show that these projectors  are derived from physical constraints; their action defines the manifold on which dynamics occur; and their mathematical structure guides the design of high-dimensional embeddings. The PrEDS formalism arises as a natural consequence of these facts, offering a general, physically principled method for modeling collective dynamics under conservation.

In the sections that follow, we apply PrEDS to a range of systems, -memristive circuits, slime mold flow networks, and adaptive mechanical lattices, -demonstrating that conservation-induced projections are not limited to theory, but manifest in a wide variety of real-world systems. Each application builds on the results derived here and explored in full in the appendices.

\subsection{Mean Field Matrix Examples}\label{appendix:meanfieldexamples}

We now illustrate the utility of the mean-field matrix approach in capturing average dynamics. Consider a one-dimensional asymmetric double-well potential: 
\begin{align}
V(x) &= -(a_0 + a_1 x + a_2  x^2 + a_3 x^3 + a_4 x^4),
\\
f(x) &=-\nabla_x V(x),
\end{align}
with $(a_0, a_1, a_2, a_3, a_4) = (0, -9.85, -40, -2, 0.395)$.
The equation for $x$ can be written as
\begin{eqnarray}
    \frac{dx}{dt}=-\partial_x V=f(x).
\end{eqnarray}
As described earlier, we now lift via PrEDS:
\begin{eqnarray}
    f(x)\rightarrow \mathbf{F}(\mathbf{\Omega} \mathbf{X}).
\end{eqnarray}
First, let us note that even in this simple example, while $f(x)$ is an actual force, e.g. is the minus the derivative of the potential, in general, the matrix function $\mathbf{F}$ in general is not. This is important, as the PrEDS embedding is \textit{not} gradient following. Despite this, it is still true by construction that the fixed points of the higher-dimensional dynamical system are still associated to the fixed points of the lower dimensional system.

\begin{figure}[ht]
\centering
\includegraphics[width=\linewidth]{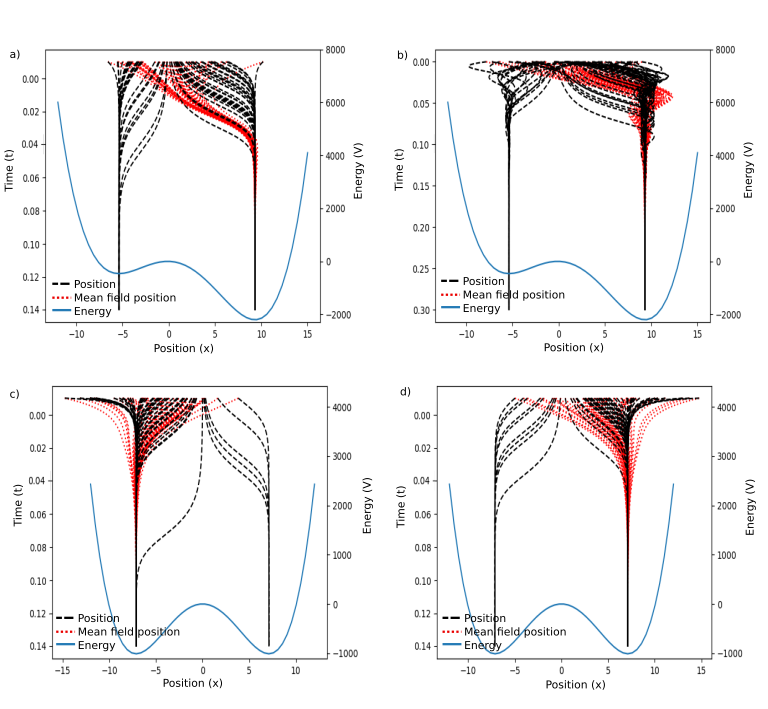}
\caption{(a) Dynamical evolution of particles in an asymmetric potential well. Individual trajectories (black) and mean-field trajectory (red) are shown. (b) Same system with momentum. (c, d) Trajectories in a symmetric double-well potential. Initial conditions are centered around the left (c) or right (d) well.}
\label{fig:Examples}
\end{figure}

We simulate two systems: with and without mean-field projection. Consider the dynamics of $N$ particles in the potential landscape given by $V(x)$, and an $N$-dimensional lifted dynamics for a single particle, where $m=1$.

In the absence of momentum:
\begin{align}
    \frac{dx}{dt} &= -\nabla_{{x}} V( x) \nonumber \\
    & \hspace{.1cm}\rightarrow 
    \frac{d\vec{X}}{dt} = \mathbf{F}(\mathbf{K}, \{\mathbf{X}\}) \cdot \vec{1} - \alpha (\mathbf{I} - \mathbf{K}) \mathbf{X} \cdot  \vec 1
    ,
    \\
    \mathbf{F}(\mathbf{K},\{\mathbf{X}\}) &=  a_1 \mathbf{K}  + 2 a_2 (\mathbf{K} \mathbf{X}) \nonumber \\
    &\hspace{0.5cm}+ 3 a_3 (\mathbf{K} \mathbf{X})^{2} + 4 a_4 (\mathbf{K} \mathbf{X})^{3},
    \label{eqn:examplewoMomentum}
\end{align}
with $\alpha = 1$. Here, $\mathbf{X}$ is the diagonal matrix of $\vec{X}$, $\{\vec{X}\}\in \mathbb{R}^{N\times 1}$, $\vec{X}=\vec{x}^{\, T}$,   $\mathbf{K}=\mathbf{I}$ in the case of $N$-independent particles, the original un-lifted dynamics and $\mathbf{K}=\mathbf{\Omega}$ in the PrEDS model.

For $N$ independent particles with momentum $\vec{p}$:
\begin{align}
    \frac{d\vec{x}}{dt} &= \frac{\vec p}{m} , \\
    \frac{d\vec{p}}{dt} &= \vec f(x) - \xi~\frac{\vec{p}}{m}, 
    \end{align}
which leads to 
    \begin{align}
    \frac{d\vec{P}_i}{dt} &= \mathbf{\Omega} \left(\mathbf{F}_i(\mathbf{\Omega},\{\mathbf{X}_i\}) - \xi \mathbf{P}_i \right) \cdot \vec{1} - \alpha (\mathbf{I} -\frac{1}{m} \mathbf{\Omega}) \vec{P}_i, \\
    \frac{d\vec{X}_i}{dt} &= \mathbf{\Omega} \frac{\vec{P}_i}{m} \cdot \vec{1} - \alpha (\mathbf{I} - \mathbf{\Omega}) \vec{X_i},
\end{align}
where $(\alpha, M, m, \xi) = (30, 0.1, 0.1, 1)$.

We numerically integrate the equations and display results in Figure~\ref{fig:Examples}. Plots (a) and (b) show trajectories in the asymmetric potential with and without momentum. Black lines represent individual particle trajectories; red lines show the mean-field trajectory, which converges to the global minimum.

We also analyze a symmetric double-well potential:
\begin{equation}
V( x) = -(a_0 + a_2  x^2  + a_4 x^4 ),
\end{equation}
with $(a_0,  a_2,  a_4) = (0,  -40,  0.395)$.
Trajectories are shown in (c) and (d) for initial positions centered around the left and right wells, respectively. Mean-field trajectories settle in the potential well closest to the average initial condition.

\section{Projection-Based Dynamics in Physical and Biological Systems}

The Projective Embedding of Dynamical Systems (PrEDS) framework is inspired by and validated through its application to a range of physical and biological systems that exhibit self-organization and adaptation under local conservation laws. In this section, we discuss three representative examples: electrical circuits with memory, adaptive flow networks inspired by slime molds, and elastic spring networks. These systems are chosen because they are both well-studied in the literature and serve as archetypes for computation, optimization, and collective dynamics.

In each example above, local conservation laws define a constrained subspace in which the system evolves. These constraints naturally give rise to projection operators, -constructed from the underlying network topology, -that govern the effective dynamics. This subspace forms a \textit{dynamical manifold}, and the system’s adaptation, memory, or optimization is confined to this manifold.

Within the PrEDS framework, we interpret these systems as higher-dimensional spaces embeddings where their dynamics  governed by projector-induced constraints. The resulting formalism unifies resistive learning in memristive circuits, flow adaptation in slime molds, and mechanical response in elastic lattices under a shared mathematical structure.

We emphasize that these results are not just heuristic. Detailed derivations and simulations are provided in Appendices ~\ref{appendix:projoperators}, ~\ref{appendix:systems} and ~\ref{appendix:BioPhysSystems}.

\subsection{Dissipative Networks}

In systems satisfying Onsager reciprocal relations, the dissipation function is given by:
\begin{equation}
     2\mathcal{D} = \vec{i}^T \vec{v},
\end{equation}
where $\mathcal{D}$ is a dissipation function \cite{Meixner_JMathPhys_1963,Meixner_1973}. For systems obeying Kirchhoff's laws and Onsager symmetry, the dynamics are governed by these projection operators.
As derived in the Appendix~\ref{appendix:nettherm}, the power dissipation defined above is then given by:
\begin{align}
    2\mathcal{D} &=  \vec{i}^T \boldsymbol{\Omega}_A \vec{v}_\text{source}, 
    \\ 2\mathcal{D} & = \vec{i}^T_\text{source}\boldsymbol{\Omega}_B\vec{v}.
    \label{eqn:Dissipations}
\end{align}
for voltage and current biased circuits, respectively.
In the Appendix, we present an alternative derivation of the dissipation in LRC circuits.


We can also express dissipation using oblique projectors. The current in a biased network is
\begin{align}
    \vec{i} &=  A^T (ARA^T)^{-1} AR \vec{i}_\text{source} = \Omega_{A/R} \, \vec{i}_\text{source}.
    \label{eqn:R_Loop_kineticMatrix}
\end{align}
A similar expression holds in the node representation:
\begin{align}
    \vec{v} =  B^T (BR^{-1}B^T)^{-1} B R^{-1}\vec{v}_\text{source} = \Omega_{B/R^{-1}} \, \vec{v}_\text{source}.
\end{align}

These flux–forcing relations, expressed through oblique projection operators, will be central to our analysis. The operators $\Omega_{B/R^{-1}}$ and $\Omega_{A/R}$ can also be interpreted as kinetic coefficients (see Appendix).

\subsection{Resistive and Memristive Circuits}

Electrical circuits provide one of the clearest examples of conservation-constrained dynamics. In particular, Kirchhoff’s Current Law (KCL) and Voltage Law (KVL) define linear constraints on the system's currents and voltages, respectively. These constraints give rise to natural projection operators on the space of possible configurations. The derivations are presented in Appendix~\ref{appendix:memristors}, but the methods are now standard as they have been derived in a series of papers \cite{Caravelli_PRE_2017,CARAVELLI2023133747}.

In a resistive circuit with voltage and current sources, the native dynamics are described by:
\begin{subequations}
\begin{align}
\vec{i} &= G \vec{v} + \vec{j}_\text{ext}, \\
\vec{v} &= R \vec{i} + \vec{s},
\end{align}
\end{subequations}
where $G$ and $R$ are the conductance and resistance matrices, and $\vec{j}_\text{ext}$ and $\vec{s}$ represent external sources. These relations, when combined with conservation laws, yield projector-based forms (see Appendix).

Memristors extend this model by adding memory: their resistance depends on an internal variable $x_i(t)$, $x\in [0,1]$, that evolves in time. A simple memristor model follows:
\begin{subequations}
\begin{align}
R(x) &= R_\text{on} x + R_\text{off}(1 - x), \\
\frac{dx}{dt} &= \frac{R_\text{off}}{\beta} i(t) - \alpha x(t),
\end{align}
\end{subequations}
where $\beta$ is an inverse learning rate, and $\alpha$ a decay constant. At the network level, the internal states evolve under projected driving:
\begin{equation}
\frac{d\vec{x}}{dt} = \frac{1}{\beta}(I - \chi \boldsymbol{\Omega}_A X)^{-1} \boldsymbol{\Omega}_A \vec{S} - \alpha \vec{x},
\end{equation}
with $\boldsymbol{\Omega}_A$ derived from the cycle matrix $A$. This reveals that the topology of the circuit constrains learning, and the dynamics are confined to a manifold defined by $\boldsymbol{\Omega}_A$.
In the projected PrEDS form:
\begin{eqnarray}
    \frac{d\vec{x}}{dt} &=& \Omega_A \left( \frac{1}{\beta} \left(I - \chi \Omega_A X \right)^{-1} \Omega_A \vec{S} - \alpha \vec{x} \right) \nonumber \\
    &&- \alpha (I - \Omega_A) \vec{x}.
\end{eqnarray}

\begin{figure*}[h!]
\centering
\includegraphics[width=\linewidth]{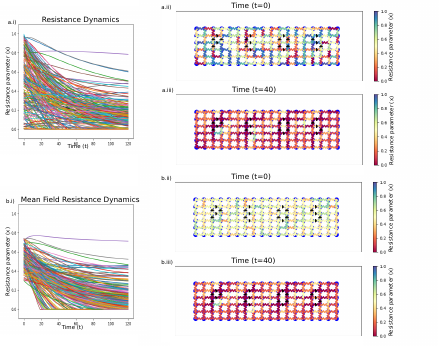}
\caption{Dynamics of a memristor network under applied bias with dimensionless parameters $(\alpha, \beta,  \chi, \alpha^\star,\text{max}|\vec{S}|) = (0.02,0.005625, 1, \frac{15}{16}, 1, 0.2)$. 
(a.i) Initial random resistance values evolve to high and intermediate states. The dashed black line indicates the average resistance, which increases due to the volatility term $\alpha$ but does not reach $R_\text{off}$. 
(a.ii) and (a.iii) show the spatial resistance profile before and after biasing; the voltage sources spelling ``PEDS", i.e., \textit{PEDS}, induce a corresponding spatial contrast in resistance. 
(b.i) Mean field dynamics show convergence to a single trajectory due to the Banality Lemma, reflecting the collective behavior of the network. 
(b.ii) and (b.iii) display convergence from a random initial resistance profile to a uniform distribution under mean field evolution.}
\label{fig:Memcircuit}
\end{figure*}

In the PrEDS framework, this system is lifted by replicating each internal variable across a higher-dimensional space and evolving them under a projector (either $\boldsymbol{\Omega}_A$ or a mean-field operator), revealing collective dynamics that converge toward globally optimal resistance patterns. This setting is also the origin of the original motivation for PrEDS, as memristive networks were shown to perform optimization and learning tasks in earlier work \cite{Caravelli_SciAdv_2021,Barrows_AdvIntSys_2025}.

Lifting the dynamics using the PrEDS formalism and instead applying the mean-field projector $\mathbf{\Omega}$, we write:
\begin{align}
 \frac{d\vec{X}_i}{dt} & = \frac{1}{\beta} \mathbf{\Omega} \left[(I - \chi \Omega_A \, \text{Diag}(\{\vec{X}\}))^{-1} \Omega_A \vec{S}\right]_i \nonumber \\
 &\qquad - \alpha \mathbf{\Omega} \vec{X}_i - \alpha^\star (\mathbf{I} - \mathbf{\Omega}) \vec{X}_i.
 \label{eqn:PrEDS_memristor}
\end{align}
This captures the mean behavior of a replicated ensemble of memristor dynamics under the same bias, consistent with the formal structure of Eq.~\eqref{eq:PrEDS_dynamics}.

In Figure \ref{fig:Memcircuit}, we demonstrate the dynamics of a dense memristor network subjected to external voltage biases. Bias nodes are chosen to achieve a desired resistance configuration, where control is exerted solely through the placement of voltage sources. This drives spatial differentiation in resistance in the network. Initial resistance values are randomly assigned and expressed via the linear memory parameter $x$. Under constant bias, the resistance values bifurcate to either a high-resistance state (driven by $\alpha$) or remain at an intermediate level, depending on their orientation relative to the voltage sources and their topological location in the circuit.

In the mean field treatment, the memristor states reflect an averaged trajectory for each memristor, and the network converges rapidly to a resistance configuration that reflects the underlying voltage generators with less local fluctuations compared to the original dynamics, consistent with previous observations \cite{Caravelli_SciAdv_2021}. Comparing the resistance profiles in Figures \ref{fig:Memcircuit}(a.iii) and (b.iii) shows that the loop projection operator retains locality: memristor updates vary based on their distance and orientation relative to the voltage sources, as previously studied in \cite{Caravelli_locality_PRE_2017}. In contrast, the mean field projection reduces spatial variation. The mean field projector does not alter the asymptotic dynamics of the network.

\subsection{Adaptive Flow Networks and Slime Molds}
\label{sec:slimemoldsummary}

The slime mold \textit{Physarum polycephalum} can be modeled as a flow-preserving network \cite{Gale2015}. Its body forms tubular structures that dynamically adapt, transporting cytoplasmic material between the main body and external food sources, effectively solving a shortest path problem \cite{Zhange_SWJ_2014}. Tubes not connected to resources tend to retract, while redundant paths diminish over time. This adaptive behavior, driven by internal feedback and local flow, is well documented \cite{Alim_PNAS_2017, Boussard2021-zx, UEDA1986486}.

We now extend the PrEDS framework to a biological flow network. Inspired by the memristor circuit, we consider a system with similar conservation laws and flux/forcing structure.

The slime mold \textit{Physarum polycephalum} can be modeled as a flow-preserving network. Its body forms tubular structures that dynamically adapt, transporting cytoplasmic material between the main body and external food sources. Tubes not connected to resources tend to retract, while redundant paths diminish over time. This adaptive behavior—driven by internal feedback and local flow—is well documented \cite{Alim_PNAS_2017, Boussard2021-zx, UEDA1986486}.

Because \textit{Physarum} preserves mass flow and responds to pressure gradients, its structure is effectively governed by local conservation laws. Cytoplasmic flow is driven by pressure differentials generated at network nodes. Although chemotaxis and signaling are involved, we neglect those in the following model and focus on fluid dynamics.

The basic flux-forcing relationship follows Poiseuille flow:
\begin{equation}
Q_{ij} = \frac{\pi \eta}{128} \frac{D^4_{ij}}{L_{ij}} (p_i - p_j),
\label{eqn:poiseuille_flow}
\end{equation}
where $p_i$ and $p_j$ are pressures at nodes $i$ and $j$, $D_{ij}$ is the diameter of the tube between them, $L_{ij}$ is the length, and $\eta$ is the fluid viscosity. Conductance is therefore proportional to $D^4_{ij}/L_{ij}$.

Flow is conserved across the network according to:
\begin{equation}
\sum_i Q_{ij} = P_0 (\delta_{bj} - \delta_{Sj}),
\label{eqn:delta_flow}
\end{equation}
where $b$ and $S$ label the body and sink (food source) nodes, and $P_0$ is the total incoming or outgoing flow. This structure mirrors Kirchhoff’s current law, allowing an interpretation in terms of effective “currents” and “voltages”.

With these relations established, we can identify the appropriate incidence matrix and projection operators to apply the PrEDS framework to this biological system.

We now apply Kirchhoff's laws to describe flow networks in the context of \textit{Physarum polycephalum}, using a graph-theoretic framework. Let $G$ be the graph representing the tubular network, and let $\vec{Q}$ be the vector of edge flows. Kirchhoff’s current law (KCL) for this system reads:
\begin{equation}
\sum_{\beta} B \vec{Q} = 0,
\end{equation}
where $B$ is the incidence matrix of the directed graph $G$. The flow through edge $(i,j)$ follows Poiseuille’s law, which is analogous to Ohm's law:
\begin{equation}
Q_{ij} = g_{ij} (p_i - p_j),
\end{equation}
with $g_{ij} = \frac{\pi \eta}{128} \frac{D_{ij}^4}{L_{ij}}$ the effective conductance, $D_{ij}$ the tube diameter, and $L_{ij}$ its length.

To include pressure sources and sinks, we introduce a virtual ground node, indexed as $0$, and define pressures relative to this node. For each node $i$, we add an edge $(i,0)$ to a new graph $G' = G \cup G_e$, where $G_e$ contains the added source/sink edges. The direction of each new edge depends on the sign of $p_i - p_0$: if $p_i > p_0$, it is directed from $0$ to $i$, and vice versa.

The pressure drop across any edge $\beta$ in the extended graph $G'$ can then be written as:
\begin{equation}
\Delta V_{\beta} =
\begin{cases}
\frac{Q_{\beta}}{g_{\beta}} & \beta \in G \\
p_i - p_0 & \beta \in G_e
\end{cases}
= \frac{Q_{\beta}}{g_{ij}} \delta_{\beta, G} + \Delta V_{\beta} \delta_{\beta, G_e}.
\end{equation}
This formulation allows us to treat pressure sources and sinks consistently, and all edges now participate in a complete network governed by Kirchhoff-type constraints.

Tellegen's theorem enforces potential conservation around cycles:
\begin{equation}
\sum_{\beta} A_{l\beta} \vec{V}_{\beta} = 0,
\end{equation}
where $A_{l\beta}$ is the fundamental cycle matrix of $G'$. The ground node is removed when computing reduced incidence and loop matrices. We also include a high-conductance $\tilde{g}$ in series with the pressure sources to maintain linearity, and take the limit $\tilde{g} \to \infty$ at the end of the calculation.

With this structure, the flow network becomes mathematically equivalent to a resistive network, with $r_{ij} \propto \frac{L_{ij}}{D_{ij}^4}$. Since the geometry of the network adapts dynamically, we introduce a memory parameter $x_{ij} \in [0, 1]$ to scale between high- and low-resistance states. We consider two adaptation regimes:

\begin{align}
R_{ij}(x) = \begin{cases}
\frac{1}{\eta'} \frac{1}{d_0^4} (L_{\text{min}} x + L_{\text{max}} (1 - x)) & \text{(dynamic length)} \\
\frac{1}{\eta'} l_0 (D_{\text{max}}^{-4} x + D_{\text{min}}^{-4} (1 - x)) & \text{(dynamic diameter)}
\end{cases}.
\end{align}

In these models, $x = 0$ corresponds to maximum resistance (e.g., elongated or narrow tubes), and $x = 1$ to minimum resistance (e.g., short or wide tubes).

\textit{Physarum} adapts its tubular structure in response to flow, captured by:
\begin{equation}
\frac{dx_{ij}}{dt} = f(Q_{ij}) - \kappa x_{ij},
\label{eqn:flow_independent}
\end{equation}
with $\kappa$ a decay parameter. We adopt the simple choice $f(Q_{ij}) = |Q_{ij}|$, as in \cite{Tero_Science_2010}. Using circuit theory, we express the flow in terms of pressure sources:
\begin{align}
\vec{Q} &= -A^T (A R A^T)^{-1} A \Delta \vec{V}_\text{source} \\
&= -\frac{1}{R_{\text{off}}} (I - \chi \Omega_A X)^{-1} \Omega_A \Delta \vec{V}_\text{source},
\label{eqn:flow_Lchanging}
\end{align}
where $\Roff = \frac{1}{\eta'} \frac{L_{\text{max}}}{D_{\text{min}}^4}$ and $\chi$ is a scaling factor from the resistance parametrization.

The nonlinear dependence of the dynamics on $|\vec{Q}|$ prevents a direct Taylor expansion. However, we can apply the PrEDS framework with mean field projection:
\begin{widetext}
\begin{align}
\frac{d\vec{x}}{dt} &= -\kappa \vec{x} + 
\begin{cases}
-\frac{1}{\Roff} ( (I - \chi \Omega_A X)^{-1} \Omega_A \Delta \vec{V}_\text{source} )_i & Q(x)_i > 0 \\
+\frac{1}{\Roff} ( (I - \chi \Omega_A X)^{-1} \Omega_A \Delta \vec{V}_\text{source} )_i & Q(x)_i \leq 0
\end{cases}, \\
\frac{d\vec{X}}{dt} &= -\kappa \mathbf{\Omega} \vec{X} - \alpha^\star (\mathbf{I} - \mathbf{\Omega}) \vec{X} + \frac{1}{\Roff}
\begin{cases}
\mathbf{\Omega} ( - (I - \chi \Omega_A \text{Diag}(\vec{X}))^{-1} \Omega_A \Delta \vec{V}_\text{source} )_i & Q(\vec{X})_i > 0 \\
\mathbf{\Omega} ( + (I - \chi \Omega_A \text{Diag}(\vec{X}))^{-1} \Omega_A \Delta \vec{V}_\text{source} )_i & Q(\vec{X})_i \leq 0
\end{cases}.
\end{align}
\end{widetext}

For clarity and simplicity, this can also be written as:
\begin{align}
\frac{d\vec{X}}{dt} &= -\kappa \mathbf{\Omega} \vec{X} + \frac{1}{\Roff} \mathbf{\Omega} \left| (I - \chi \Omega_A \text{Diag}(\vec{X}))^{-1} \Omega_A \Delta \vec{V}_\text{source} \right| 
\nonumber \\ &\qquad\qquad
- \alpha^\star (\mathbf{I} - \mathbf{\Omega}) \vec{X}.
\end{align}
This formulation provides a compact expression for the evolution of \textit{Physarum}'s tubular network under the PrEDS framework, incorporating both local adaptation and global averaging via the mean field projection.

\subsection{Elastic Spring Networks}
\label{sec:spring-networks}

Another example of an adaptive system considered in the literature for learning is a system of spring networks.
Spring networks serve as prototypical models for adaptive mechanical systems and have been widely used to explore principles of distributed computation in soft materials and mechanical learning systems \cite{Stern_PRX_2021, patil2023selflearningmechanicalcircuits}. The detailed derivations are presented in Appendix~\ref{appendix:springnetworks}. We show that also in this case, spring networks can be written in terms of projector operators,  and then apply the PrEDS framework to such systems and show how local adaptation rules, constrained by mechanical conservation laws, give rise to emergent structure.

Each edge of the network connects a pair of nodes via a Hookean spring with stiffness $k_{ij}$, equilibrium extension $\vec{X}_{ij}^{\mathrm{eq}}$, and dynamic displacement vector $\vec{x}_{ij}$. At mechanical equilibrium, the total force at each node must vanish. This local conservation law is expressed as:
\begin{equation}
\sum_j \vec{F}_{ij} = 0,
\end{equation}
where the force along spring $(i,j)$ is $\vec{F}_{ij} = -k_{ij}(\vec{x}_{ij} - \vec{X}_{ij}^\text{eq})$.

To model adaptation, we introduce a memory variable $z_{ij}(t) \in [0,1]$ that determines the spring stiffness as 
\begin{equation}
k_{ij}(t) = k_\text{min} + (k_\text{max} - k_\text{min}) z_{ij}(t),
\end{equation}
with $\xi = (k_\text{max} - k_\text{min}) / k_\text{min}$ controlling the dynamic range. The variable $z_{ij}$ evolves under an energy-based adaptation rule:
\begin{equation}
\frac{d z_{ij}}{dt} = \alpha z_{ij} - \frac{1}{\beta} \Delta \ell_{ij}^2,
\end{equation}
where $\Delta \ell_{ij}$ is the deviation of the spring from its equilibrium length. The energy term $\Delta \ell_{ij}^2$ serves as the conjugate force driving stiffness change.

As detailed in Appendix~\ref{appendix:springnetworks}, we formulate the mechanical response using a node-based projection operator $\boldsymbol{\Omega}_B = B^T (B B^T)^{-1} B$, derived from the incidence matrix $B$. The total spring displacement vector $\Delta \vec{x}$ (decomposed into $x$ and $y$ components) satisfies:
\begin{equation}
\Delta \vec{x}^\mu = - (I + \xi \boldsymbol{\Omega}_B Z)^{-1} \boldsymbol{\Omega}_B \vec{F}^\mu_\text{bias}, \quad \mu \in \{x, y\},
\end{equation}
where $Z = \text{diag}(\vec{z})$ is the stiffness memory matrix and $\vec{F}^\mu_\text{bias}$ denotes the external force projection in direction $\mu$. This projection ensures that all displacements respect the topological constraints imposed by the network, including force balance and cycle consistency.

The dynamics of the memory variable $z_{ij}$ are then governed by the total squared displacement:
\begin{equation}
\frac{d z_{ij}}{dt} = \alpha z_{ij} - \frac{1}{\beta} \left( (\Delta x_{ij})^2 + (\Delta y_{ij})^2 \right),
\end{equation}
with the projection appearing implicitly via $\Delta \vec{x}^\mu$ as defined above.

The mean-field PrEDS embedding, using a uniform projector $\mathbf{\Omega}$ over the spring memory variables, captures the averaged system behavior across replicas:
\begin{eqnarray*}
\frac{d \vec{Z}_i}{dt} &=& \alpha \mathbf{\Omega} \vec{Z}_i - \frac{1}{\beta k_\text{min}^2} \mathbf{\Omega} \left[ \left( (I + \xi \boldsymbol{\Omega}_B \text{Diag}(\{\vec{Z}\}))^{-1} \boldsymbol{\Omega}_B \vec{F}^\mu_\text{bias} \right)^2 \right]_i \nonumber \\
&&- \alpha^\star (\mathbf{I} - \mathbf{\Omega}) \vec{Z}_i.
\end{eqnarray*}
This lifted representation reveals that mean-field dynamics suppress buckling and local instability by biasing the network toward globally consistent stiffness configurations.

Simulation results (see Appendix) confirm that individual networks exhibit diverse local trajectories due to heterogeneous adaptation, while the mean-field version converges smoothly to mechanically stable states. The resulting spring configurations minimize stored elastic energy while respecting geometric and topological constraints, illustrating how constrained physical systems can compute optimized structures through local adaptation.

\section{Mapping PrEDS to a particle swarm model}

We wish now to provide a unifying concept to understand how PrEDS maps all these physical and biological systems to a certain class of self-organizing particle swarm models. Specifically, we interpret the PrEDS dynamics as an interacting particle system, which not only clarifies the structure of the governing equations, but also reveals a natural connection to heuristic optimization frameworks, particularly those inspired by swarm intelligence, and for all the systems mapped to PrEDS. In such algorithms, originally developed to model the collective behavior of social organisms, a population of agents explores a solution landscape by combining local sensing with global coordination. This idea underpins methods like Ant Colony Optimization and Particle Swarm Optimization, which have proven effective across a range of nonlinear and combinatorial problems \cite{dorigo1999ant,theraulaz2002swarm,Caravelli2015}. By recasting PrEDS as a particle swarm, we show that the emergent behavior of the system, e.g., converging toward global minima through local interactions, shares deep parallels with these optimization techniques. This interpretation positions PrEDS as a principled framework for understanding how distributed agents can collectively perform optimization and decision-making in complex environments.

We interpret eq.~\eqref{eq:PrEDS_dynamics}  with commutative map formalism under the framework of particle interactions, we establish a direct correspondence between the governing equations and the collective behavior of a multi-particle system.  To achieve this, we define particle positions $\vec{r}_1, ... \vec{r}_N$ in $\mathbb{R}^m$ where the position of the $\beta-th$ particle is $\vec{r}_{\beta} = [X_{1,\beta},X_{2,\beta},...X_{m,\beta}]^T $.  Here, We use Greek letters as subscripts to denote particle numbers and Latin letters for components of position vectors. We rewrite eq.~\ref{eq:PrEDS_dynamics} as 
\begin{align}
     \frac{d \vec{r}_{\beta}}{dt} 
     &= \frac{1}{N} \sum_{\theta} \vec{f}(\vec{r}_{\theta}) - \frac{\alpha}{N}\sum_{\theta}\vec{r}_{\beta} - \vec{r}_{\theta} \\
     &= -\frac{1}{N} \sum_{\theta}  \nabla_{\vec{r}_\theta} V(\vec{r}_\theta) - \frac{\alpha}{N}\sum_{\theta} \nabla_{\vec{r}_\beta} U_H(\vec{r}_{\beta} - \vec{r}_{\theta}).
    \label{eq:PrEDS_particle_dynamics}
\end{align}
Eq.~\ref{eq:PrEDS_particle_dynamics} demonstrates that PrEDS is equivalent to the interacting particle swarm (see Appendix~\ref{appendix:Swarm} for more details). To be more specific, the first term in eq.~\ref{eq:PrEDS_particle_dynamics} shows that particles communicate by sensing others' gradients and moving along the average gradient direction, given by $-\frac{1}{N} \sum_{\theta}  \nabla_{\vec{r}_\theta} V(\vec{r}_\theta)$. The second term is the harmonic attraction where $U_H(\vec{x}) = -\frac{1}{2}|\vec{x}|^2$. The overall dynamics is composed of the descent along the average gradient and the pairwise attraction among particles. Together, these terms define a dynamic system in which particles both descend along the average gradient and experience pairwise attraction. Intuitively, particles that fall into deeper minima can pull others in, leading to a possible equilibrium state where all particles accumulate in a deep minimum.

\subsection{Evolution of the density field for PrEDS particle swarm}
To understand the emergent collective behavior governed by the particle dynamics described in eq.~\eqref{eq:PrEDS_dynamics}, we adopt a continuum description of the system. This approach becomes increasingly accurate in the limit of a large number of particles, i.e., as $ N \rightarrow \infty $, under the mean-field approximation, which assumes weak two-point correlations between particles.

Within this continuum framework, the particle density field captures the macroscopic manifestation of individual particle interactions. Following the method in \cite{dean1996langevin}, we represent the instantaneous particle density field of the swarm with positions $ \vec{r}_1, \ldots, \vec{r}_N $ as a superposition of Dirac delta functions:
\begin{equation}
    \rho(\vec{r},t) = \sum_{\beta} \rho_{\beta}(\vec{r},t),
\end{equation}
where each particle contributes a density component defined as 
\begin{equation}    
\rho_{\beta}(\vec{r},t) \equiv \delta(\vec{r} - \vec{r}_\beta(t)).
\end{equation}
This expression characterizes the exact microscopic density field of the system at time $ t $.

To analyze the time evolution of this field, we introduce an auxiliary function $ f(\vec{r}_\beta(t)) $, which evaluates a smooth test function $ f(\vec{r}) $ at the particle's position using the delta function representation:
\begin{equation}
    f(\vec{r}_\beta(t)) = \int d\vec{r} \ \delta(\vec{r} - \vec{r}_\beta(t)) f(\vec{r}) .
\end{equation}

Taking the time derivative of this expression (see Appendix~\ref{appendix:Swarm} for derivation), we obtain:
\begin{widetext}
\begin{align}
    \frac{d}{dt}f(\vec{r}_\beta(t)) 
    &= \int d\vec{r} \ f(\vec{r})  \nabla_{\vec{r}} \cdot \left[\rho_\beta(\vec{r},t)\left(\frac{1}{N} \sum_{\theta}  \nabla_{\vec{r}_\theta} V(\vec{r}_\theta) + \frac{\alpha}{N}\sum_{\theta} \nabla_{\vec{r}} U_H(\vec{r} - \vec{r}_\theta)\right)\right].
    \label{eq:dt_f}
\end{align}
\end{widetext}

Comparing this expression with the identity
\begin{equation}
\frac{d}{dt}f(\vec{r}_\beta(t)) = \int d\vec{r} \ f(\vec{r}) \, \partial_t \rho_\beta(\vec{r},t),
\end{equation}
we can directly identify the time evolution equation for the single-particle density field $ \rho_\beta(\vec{r},t) $:
\begin{widetext}
\begin{equation}
    \partial_t \rho_\beta(\vec{r},t) = \nabla_{\vec{r}} \cdot \left[\rho_\beta(\vec{r},t)\left(\frac{1}{N} \sum_{\theta} \nabla_{\vec{r}_\theta} V(\vec{r}_\theta) + \frac{\alpha}{N} \sum_{\theta} \nabla_{\vec{r}} U_H(\vec{r} - \vec{r}_\theta)\right)\right].
\end{equation}
\end{widetext}

To obtain the evolution of the full density field $ \rho(\vec{r},t) $, we sum the above expression over all particles indexed by $ \beta $. In doing so, we replace the discrete summation over particles with integrals weighted by the density field itself. This yields the following partial differential equation for the collective particle density:
\begin{widetext}
\begin{align}
    \partial_t \rho(\vec{r},t) 
    &= \nabla_{\vec{r}} \cdot \left[ \rho(\vec{r},t) \left( \frac{1}{N} \int d\vec{y} \, \nabla_{\vec{y}} V(\vec{y}) \rho(\vec{y},t) + \frac{\alpha}{N} \nabla_{\vec{r}} \int d\vec{y} \, U_H(\vec{r} - \vec{y}) \rho(\vec{y},t) \right) \right].
    \label{eq:dt_rho}
\end{align}
\end{widetext}

This equation governs the evolution of the macroscopic density field under the influence of two contributions: (i) the interaction potential $ V $ acting on the particle positions, and (ii) a convolution term involving the interaction kernel $ U_H $, which introduces nonlocal effects. The mean-field approximation ensures that the influence of each particle is effectively averaged over the distribution, making eq.~\eqref{eq:dt_rho} a useful description for analyzing the large-scale dynamics of the system.

\subsection{Numerical Simulations of Particle Swarms}

\begin{figure*}[h!]
\centering
\includegraphics[width=1.0\textwidth]{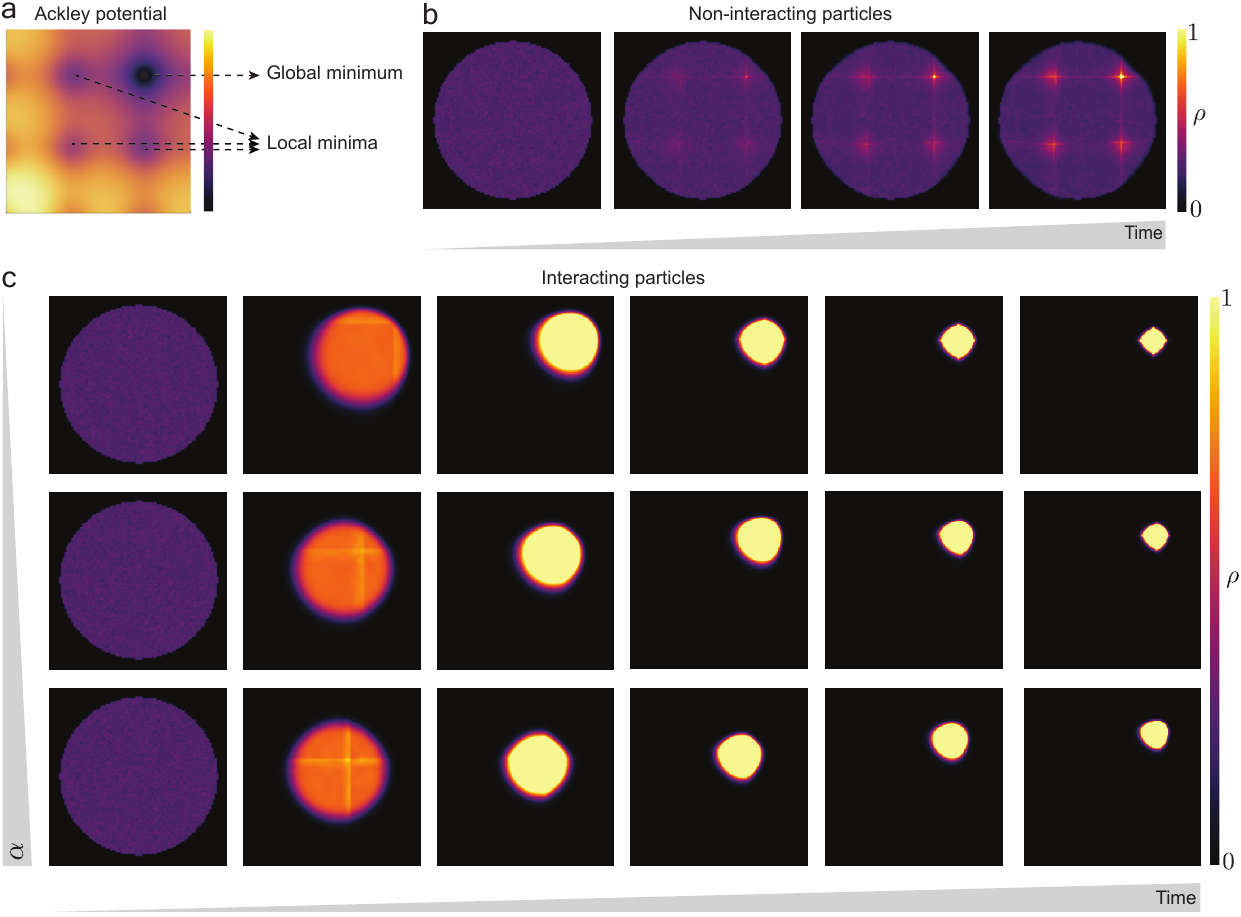}
\caption{
(a) Ackley function with three local minima and one global minimum as a test potential to study PrEDS. The Ackley function in two dimensions is given as: $V(x, y) = -20 \exp \left( -0.2 \sqrt{\frac{1}{2} ((x - 1.875)^2 + (y - 1.875)^2)} \right) - \exp \left( \frac{1}{2} (\cos(2\pi (x - 1.875)) + \cos(2\pi (y - 1.875))) \right) + 20 + e$. 
(b) Density evolution in finite-difference simulations of non-interacting particles (Eq.~\eqref{eq:dt_rho_non_int}), consecutive frames show density snapshots at intervals of $20 dt$ starting from $t=0$. (c) Density evolution in finite-difference simulations of interacting particles (Eq.~\eqref{eq:dt_rho}). Three different rows represent different values of parameter $\alpha$, with $\alpha = 1$, $5$, $10$ respectively.  For rows 1, 2, and 3, consecutive frames are at intervals of $50 dt$, $10 dt$, and $5 dt$, respectively. Initial density  for both (a) and (b) is a circular Gaussian distribution of mean $\mu = N/L_xL_y$, and standard deviation $\sigma$, where $\int \rho \, d\mathbf{x} = N$, and $L_x, L_y$ denote the side lengths of the simulation box. We use Euler time-stepping with a step size $dt$ and the first-order upwind scheme with a square-grid spatial discretization $dx, dy$ to solve Eqs.~\ref{eq:dt_rho_non_int} and \ref{eq:dt_rho}. We set $N = 1.0$, $L_x = L_y = 2.5$, $\sigma = 0.1 \mu$, $dt = 0.0001$, $dx=dy=0.025$.}
\label{fig:Swarm}
\end{figure*}

We perform finite-difference simulations of particle swarms governed by eq.~\eqref{eq:dt_rho}, using the PrEDS formalism. As a test case, we use the Ackley function, a multi-modal potential with several local minima and a single global minimum, shown in Fig.~\ref{fig:Swarm}(a). This allows us to investigate how particle swarms explore complex landscapes.

Remarkably, we observe that an initially diffuse particle density consistently aggregates and converges toward the global minimum, as shown in Fig.~\ref{fig:Swarm}(c). Furthermore, increasing the parameter $\alpha$ leads to an initial flow of density toward the system’s center of mass, followed by convergence to the global minimum. This behavior arises because $\alpha$ controls the strength of the second term in eq.~\eqref{eq:dt_rho}, which acts as a harmonic attraction toward the ensemble mean position.

These results demonstrate that PrEDS-guided particle swarms can effectively escape local minima and locate the global minimum with high reliability. To highlight the advantage of this method, we compare it against a baseline of non-interacting dynamics governed solely by gradient descent:
\begin{align}
    \partial_t \rho(\vec{r},t) 
    &= \nabla_{\vec{r}} \cdot \left[ \rho(\vec{r},t) \nabla_{\vec{r}} V(\vec{r}) \right].
    \label{eq:dt_rho_non_int}
\end{align}

Figure~\ref{fig:Swarm}(b) shows that in the non-interacting case, the steady-state density remains spread across all minima, weighted by their relative depths. In contrast, PrEDS dynamics consistently concentrate density at the global minimum. This contrast underscores the utility of PrEDS as a collective optimization mechanism superior to simple gradient descent.

\section{Discussion}

In this work, we explored the Projective Embedding of Dynamical Systems (PrEDS), a framework originally introduced in Ref.~\cite{CARAVELLI2023133747} to describe mean-field embeddings of dynamical systems governed by conservation laws beyond the context of electrical circuits. The core idea behind PrEDS is that conservation constraints, arising naturally in physical and biological networks, restrict the system’s dynamics to a structured subspace. This constraint can be encoded using projection operators, which then define the dynamical manifold on which the system evolves.

Here, we extended the scope of PrEDS beyond its original context in circuits, demonstrating its applicability to a broader class of systems where adaptation and memory occur under conservation laws. Specifically, we applied the framework to resistive and memristive circuits, adaptive flow networks inspired by Physarum polycephalum, commonly known as slime molds, and elastic spring lattices with evolving stiffness. In each case, we derived the relevant network projectors from the underlying graph structure and showed how they constrain both fast dynamics and memory evolution. We have shown that both these systems can be expressed as PrEDS.

We also clarified the physical basis of these projectors. By revisiting classical results from nonequilibrium thermodynamics, including Onsager’s reciprocal relations and entropy production, we derived orthogonal and oblique projectors from a variational principle over dissipative dynamics. These derivations, presented in Appendix~\ref{appendix:projoperators}, provide a foundation for interpreting the projectors not just as algebraic constructions but as physically meaningful operators.

A key result of this work is the identification of a formal correspondence between mean-field PrEDS dynamics and swarm-like interacting particle systems. We showed that PrEDS equations can be recast as gradient descent with harmonic coupling, and that this structure supports a continuum description of collective evolution. This reinterpretation is useful for understanding how PrEDS relates to distributed optimization and coordination strategies found in swarm intelligence.

Across all systems studied, we found that the projection operators play a dual role: they constrain the space of allowable configurations, and they shape the collective dynamics of adaptation. While the examples considered are specific, the underlying mechanism, e.g. the emergence of a dynamical manifold from local conservation laws, may be more broadly relevant.

Looking forward, the PrEDS framework offers several directions for extension. Incorporating stochastic forcing or fluctuation–dissipation relations would allow for applications in noisy, far-from-equilibrium systems. Allowing projectors to evolve over time or vary spatially could provide a path toward modeling heterogeneity in biological and engineered networks. Additionally, connections between PrEDS and methods in control, machine learning, or distributed optimization remain worth exploring.

In summary, this work clarifies the role of network-derived projection operators in shaping the dynamics of constrained adaptive systems. By extending PrEDS to multiple physical domains and identifying its relation to swarm dynamics, we contribute to a growing understanding of how conservation laws and network structure can guide complex collective behavior.
In the future, we will focus on further extending this formalism to a larger class of complex adaptive systems.

\bibliographystyle{ieeetr}
\bibliography{PEDSbib}

\clearpage
\onecolumngrid

\appendix

\section{Graph theory}
\label{appendix:projoperators}

\subsection{Example}
\begin{figure}[ht]
\centering
\includegraphics[width=.2\textwidth]{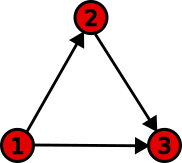}
\caption{ A simple graph with three edges connecting three nodes without an oriented cycle.}
\label{fig:SimpleTriangle}
\end{figure}

 As a simple example, let us work with the triangle graph defined by,
\begin{align}
V = \{1, 2, 3\}, \quad E = \{(1,2), (2,3), (1,3)\}
\end{align}
A representation of this graph is shown in Figure \ref{fig:SimpleTriangle}.
Note that the edge between 1 and 3 is not cyclical. 

For our triangle graph, this is,
\begin{equation}
B = \bordermatrix{
  & e_{12} & e_{23} & e_{13} \cr
  n_1 &1  &  0 &  1 \cr
n_2 &-1 &  1 &  0 \cr
n_3 &0  & -1 & -1 \cr
}
\end{equation}
$e_{ij}$ are edges linking nodes $i$ and $j$ (edges are indexed with superscripts, $e^k$), and $n_i$ are indexed nodes. 

 The triangle graph has two possible cycles, $1\to 2 \to 3\to 1$ or $1 \to 3 \to 2 \to 1$.  We can include both for now,
\begin{equation}
A =  \bordermatrix{
& e_{12}& e_{23} & e_{13} \cr
\text{cycle}_1&1 & 1 & -1 \cr
\text{cycle}_2&-1 & -1 & 1 }
\end{equation}
These are not independent, and we must eliminate one cycle to solve for a voltage configuration, generating a reduced cycle matrix.
\subsection{ Fundamental current cycle and potential at nodes}
A current $\vec{i}$ may then be expressed by the \emph{fundamental loop currents} $\vec q$ as a weighted sum of the fundamental cycles. Writing a \emph{reduced cycle matrix} $A$ in terms of the fundamental loops, we have
\begin{align}
\vec{i} = A^t \vec{q} .\end{align}
It follows that
\begin{align}
\vec{q} = (A A^t)^{-1}  A \vec{i} .\end{align}
Given a spanning tree $T$, we can find unique paths from any arbitrary initial node, the root, and any other node $k$. We can sum the voltages at the nodes along this path; if an edge is oriented along the walk from root to node, the voltage value of that edge is added; if the walk and edge are unaligned, the voltage is subtracted. We define this walk $\phi_k$, and the voltage configuration can be written in terms of this fundamental walk:
\begin{align}
\vec{v} = B^t \vec{\phi},\end{align}
which reproduces the relation between the voltage and potential above.
These identities will be useful to derive the results below.

\subsection{Projection Operators for Networks}
It is apparent that $\boldsymbol{\Omega}_B$ and $\boldsymbol{\Omega}_A$ are projection operators,
\begin{align}
    \boldsymbol{\Omega}_B\boldsymbol{\Omega}_B &=B^t(B B^t)^{-1}BB^t(B B^t)^{-1}B \nonumber\\
    &=\boldsymbol{\Omega}_B ,
    \\
    \boldsymbol{\Omega}_A\boldsymbol{\Omega}_A &=A^t(A A^t)^{-1}AA^t(A A^t)^{-1}A \nonumber\\
    &=\boldsymbol{\Omega}_A .
\end{align}
These are two distinct projection operators, the node and loop projection operators, respectively. These are orthogonal projections, as $AB^t=0$ via Tellegen's theorem, thus
\begin{align}
    \boldsymbol{\Omega}_A\boldsymbol{\Omega}_B &=A^t(A A^t)^{-1}AB^t(B B^t)^{-1}B \nonumber\\
    &=0\\
    \boldsymbol{\Omega}_A &=I-\boldsymbol{\Omega}_B
\end{align}

These projection operators can now be leveraged to study dynamical systems. 
\subsection{Power dissipation with $\boldsymbol{\Omega}_B$}
We start with the form of power dissipation given in the main text, $2\mathcal{D}=\vec{i}^T R \vec{i}=\vec{v}^t R^{-1} \vec{v}$, and use $G=R^{-1}$ Using Kirchhoff's current law, we can write 
\begin{align}
0&=B\vec{i} \nonumber \\
&= BR^{-1}\vec{v}\pm B\vec{i}_\text{source}
\end{align}
Here  $\vec{i}_\text{source}$ are current sources in the network, $\pm$ indicates the freedom in attaching current sources relative to the edge orientation and current along an edge depends on the potential drop across the edge and local current sources.
We note the voltage drop along edges in the network can be written in terms of the potential on the nodes, ($\vec{\phi}$), as $\vec{v}=B^t\vec{\phi}$. By taking the left pseudoinverse we have $(BB^t)^{-1}B\vec{v}=\vec{\phi}$. Now we can rewrite the dissipation as 
\begin{align}
    2\mathcal{D} &=\vec{v}^T R^{-1} B^T \vec{\phi} \nonumber
    \\ &= (B\vec{i}+B\vec{i}_\text{source})^T\vec{\phi}
    \nonumber
    \\
    &= (B\vec{i}+B\vec{i}_\text{source})^T (BB^t)^{-1}B\vec{v}
    \nonumber \\
 &= (\vec{i}+\vec{i}_\text{source})^T\boldsymbol{\Omega}_B\vec{v}
 \nonumber \\
  &= \vec{i}^T_\text{source}\boldsymbol{\Omega}_B\vec{v}.
  \label{eqn:NodeDissipation}
\end{align}
Thus, we arrive at a form of the power dissipation in terms of the projection operator, $\boldsymbol{\Omega}_B$.

\subsection{Entropy production}
From eq.~\eqref{eqn:R_Loop_kineticMatrix}, flux and force are related by a constitutive law, which we can write in terms of a kinetic matrix $L=A^T(ARA^T)^{-1}A$. 
For an isothermal conductor the entropy generated when charge $dq$ crosses a potential drop $v$ is $dS=(v/T)dq$. For a network, 
\begin{align}
\dot{S} &=\frac{1}{T}\vec{v}^T L\vec{v}_\text{source} 
\nonumber\\
&=\frac{1}{T}\vec{v}^T L\vec{v} .
\end{align}
The matrix $AA^T$ is in the form of a Laplacian matrix and positive semi-definite. As $R$ is a completely positive diagonal matrix, then $L=R^{-1/2}(R^{1/2}A^T)(ARA^T)(AR^{1/2})R^{-1/2}$, is positive semi-definite. Therefore $\dot{S}\geq 0$, and the $A^T(ARA^T)^{-1}A$ is the kinetic matrix relevant for driven circuits.

\section{Network Dynamics, Projection Operators, and PrEDS}
\label{appendix:systems}

Many physical, biological, and engineered systems can be represented as graphs, a structure that is particularly amenable to analysis. In systems governed by local conservation laws, such as those expressed in Kirchhoff's circuit laws, projection operators naturally emerge. In this section, we show that these projection operators arise from considerations of power dissipation in dissipative networks. Identifying them provides a unified formalism for describing the dynamics of dissipative systems that obey flux–forcing relationships and may exhibit memory effects that adapt to flux or forcing over time. Furthermore, the dynamics of interacting networks can be effectively modeled using mean-field projection operators, which reduce the complexity of the system by projecting onto a lower-dimensional subspace that captures the system’s asymptotic behavior.

\subsection{Derivation of Projection Operators via Network Thermodynamics}\label{appendix:nettherm}

In this section, we introduce a physically motivated derivation of projection operators in networks, less abstract and more intuitive than previous mathematical treatments \cite{Caravelli_PRE_2017}. This approach also provides a natural interpretation for the projector operators $\boldsymbol{\Omega}_B$ and $\boldsymbol{\Omega}_A$.

We consider a network of resistive elements with voltage sources in series, although the formalism can be generalized to include inductors and capacitors. Let $u_l = \sum_i A_{li} v_i$ denote the loop voltage, where $A$ is the cycle matrix. In the mesh formalism, we define matrices $\tilde{C}$, $\tilde{R}$, and $\tilde{L}$ representing the mesh-level capacitances, resistances, and inductances, respectively. The diagonal entries of $\tilde{C}$ give the total capacitance within a loop, and the off-diagonal entries represent shared capacitance between loops. Similar definitions apply to $\tilde{R}$ and $\tilde{L}$.

Let $Q_i$ and $\dot{Q}_i$ denote the charge and current on edge $i$, with orientation chosen arbitrarily. The total charge in loop $l$ is given by $q_l = \sum_i A_{li} Q_i$. The dynamics of the network then satisfy:
\begin{equation}
    u_i = \sum_k \left( \tilde{\Gamma}_{ik} q_k + \tilde{R}_{ik} \dot{q}_k + \tilde{L}_{ik} \ddot{q}_k \right),
    \label{eq:dissrep}
\end{equation}
where $\tilde{\Gamma} = \tilde{C}^{-1}$ is the inverse of the mesh capacitance matrix. These equations can be derived from the Lagrangian:
\begin{equation}
    \mathcal{L}(q_i, \dot{q}_i) = -\frac{1}{2} \vec{q}^T \tilde{\Gamma} \vec{q} + \frac{1}{2} \dot{\vec{q}}^T \tilde{L} \dot{\vec{q}} + \vec{u} \cdot \vec{q},
\end{equation}
together with a dissipation function $\mathcal{D}(\dot{q}_i) = \frac{1}{2} \dot{\vec{q}}^T \tilde{R} \dot{\vec{q}}$.

From this, the equations of motion for a dissipative LC circuit are given by:
\begin{equation}
    \frac{\partial \mathcal{L}}{\partial q_l} - \frac{d}{dt} \frac{\partial \mathcal{L}}{\partial \dot{q}_l} = \frac{\partial \mathcal{D}}{\partial \dot{q}_l}.
\end{equation}

In nonequilibrium thermodynamics, the thermodynamic affinities are defined as:
\begin{equation}
    \mathcal{A}_l = \frac{\partial \mathcal{L}}{\partial q_l} - \frac{d}{dt} \frac{\partial \mathcal{L}}{\partial \dot{q}_l},
\end{equation}
and the dissipation can be expressed as:
\begin{equation}
    \mathcal{D}(\dot{q}) = \sum_l \dot{q}_l \mathcal{A}_l.
\end{equation}

Let us now formulate this in terms of the edge-wise matrices. Define diagonal matrices $C$, $R$, and $L$ over all branches $\mathcal{B}$ such that $C_{bb} = C_b$ if a capacitor exists on branch $b$, and zero otherwise (similarly for $R$ and $L$). Then:
\begin{equation}
\tilde{R} = ARA^T, \quad \tilde{C} = ACA^T, \quad \tilde{L} = ALA^T,
\end{equation}
where $A$ is the loop matrix. We focus here on purely resistive networks.

Using the identity:
\begin{equation}
    \dot{\vec{q}} = (AA^T)^{-1} A \vec{i},
    \label{eqn:CycleCurrent_q}
\end{equation}
we substitute into the dissipation function:
\begin{align}
    2\mathcal{D} &= \left( (AA^T)^{-1} A\vec{i} \right)^T ARA^T (AA^T)^{-1} A \vec{i} \nonumber \\
    &= \vec{i}^T A^T (AA^T)^{-1} ARA^T (AA^T)^{-1} A \vec{i} \nonumber \\
    &= \vec{i}^T \boldsymbol{\Omega}_A R \boldsymbol{\Omega}_A \vec{i}.
\end{align}
Here, the projection operator $\boldsymbol{\Omega}_A = A^T (AA^T)^{-1} A$ arises naturally from the dissipation expression. 

Entropy production is a hallmark of nonequilibrium thermodynamics \cite{PhysRevE.110.034313}. In passive systems (e.g., resistors), dissipation directly reflects entropy production. In active or memory-bearing systems (e.g., memristors), energy input leads to irreversible changes such as heat generation or memory updates \cite{OSTER_Nat_1971}. This generalizes Tellegen’s theorem, which asserts that total power in a closed circuit sums to zero.

In systems satisfying Onsager reciprocal relations, the dissipation function (or entropy production) is given by:
\begin{equation}
    \vec{i}^T \vec{v} = \sigma T = 2\mathcal{D},
\end{equation}
where $\sigma$ is the entropy production and $T$ the temperature \cite{Meixner_JMathPhys_1963,Meixner_1973}. For systems obeying Kirchhoff's laws and Onsager symmetry, the dynamics are governed by these projection operators.
To make this explicit, we rewrite Kirchhoff’s voltage law as:
\begin{align}
    0 &= A\vec{v} \nonumber \\
      &= AR \vec{i} \pm A\vec{v}_\text{source},
\end{align}
where $\vec{v}_\text{source}$ represents external driving and $\pm$ indicates freedom in attaching the voltage sources relative to edge direction.

From the power dissipation defined above is then given by:
\begin{align}
    2\mathcal{D} &= \vec{i}_c^T A R\vec{i} = \vec{i}^T \boldsymbol{\Omega}_A (\vec{v} + \vec{v}_\text{source}) = \vec{i}^T \boldsymbol{\Omega}_A \vec{v}_\text{source}.
    \label{eqn:LoopDissipation}
\end{align}

An analogous expression holds in terms of node projection operators:
\begin{align}
    2\mathcal{D} = \vec{i}^T_\text{source}\boldsymbol{\Omega}_B\vec{v}.
\end{align}

The operators $\boldsymbol{\Omega}_A$ and $\boldsymbol{\Omega}_B$ are dual and orthogonal: $\boldsymbol{\Omega}_A \boldsymbol{\Omega}_B = 0$ and $I - \boldsymbol{\Omega}_B = \boldsymbol{\Omega}_A$.

We can also express dissipation using oblique projectors. From $R^{-1/2} \vec{v} = R^{1/2} A^T \vec{i}_c \mp R^{1/2} \vec{i}_\text{source}$, we derive:
\begin{align}
    \vec{i} &= A^T (ARA^T)^{-1} A \vec{v} \pm A^T (ARA^T)^{-1} A  R\vec{i}_\text{source} = \pm A^T (ARA^T)^{-1} AR \vec{i}_\text{source}.
    \label{eqn:R_Loop_kineticMatrix}
\end{align}

A similar expression holds in the node representation:
\begin{align}
    \vec{v} = \pm B^T (BR^{-1}B^T)^{-1} B R^{-1}\vec{v}_\text{source}.
\end{align}


In driven systems, entropy production has a minimal form that is invariant under transformations of flux and forcing \cite{Agren_JPhaseEqui_2022}. The projection operators $\boldsymbol{\Omega}_A$ and $\boldsymbol{\Omega}_B$ implement such invariant transformations, preserving reciprocity in mesh and node representations. For example, transforming $\vec{i} \rightarrow B\vec{i}$ induces a corresponding transformation on $\vec{v}$, and this is captured by the action of $\boldsymbol{\Omega}_B$, constructed via the Moore–Penrose pseudoinverse.

\subsection{Oblique projection operators}
We arrive at the form $(ARA^t)^{-1}AR\vec{i}_\text{source}=\vec{i}_c$ in the previous section by noting $R^{-1/2}\vec{v}=R^{1/2}A^t\vec{i}_c - R^{-1/2}\vec{i}_\text{source}$, we act with the left pseudoinverse $(R^{1/2}A^t)^+$ to arrive at the form $(ARA^t)^{-1}AR\vec{i}_\text{source}=\vec{i}_c$. 

Following the same procedure above and using conductivities $G=R^{-1}$, we fist note that $G^{-1/2}\vec{i}= G^{1/2}B^t\vec{\phi} - G^{1/2}\vec{v}_\text{source}$, acting with the left pseudoinverse $(G^{1/2}B^t)^+$, we arrive at the form $(BGB^t)^{-1}BG\vec{v}_\text{source}=\vec{\phi}$. 

Now, from the first line in eq.~\eqref{eqn:NodeDissipation}, we can write the dissipation as
\begin{align}
      2 \mathcal{D} &=\vec{v}^T G B^T(BGB^t)^{-1}BG\vec{v}_\text{source} \nonumber
    \\&=\vec{v}^T G \Omega_{B/G}\vec{v}_\text{source} 
    \label{eqn:G_NodeDissipation}
\end{align}
where $\Omega_{B/G}$ is an oblique projection operator. From the second line in eq. \eqref{eqn:G_NodeDissipation}, we can write (generalizing to $\pm\vec{v}_\text{source}$ to account for freedom in how sources are attached):
\begin{align}
\vec{v} &=\mp \Omega_{B/G} \vec{v}_\text{source} \nonumber \\ 
 &= \mp B^T(BGB^T)^{-1}B\vec{i}_\text{source} .
\label{eqn:G_Node_kineticMatrix}
\end{align}
This reproduces the form given above.

\subsection{Brief Review of PrEDS}\label{appendix:PrEDSformulation}

Embedding dynamical systems via projection operators provides a powerful framework for analyzing complex systems that satisfy conservation laws. The asymptotic dynamics are confined to a relevant subspace defined by these operators. In the PrEDS method, the original dynamical variable $\vec{x} \in \mathbb{R}^m$ is embedded in a high-dimensional space $\mathbb{R}^{N \times m}$, forming a set of variables $\{\vec{X}_1, \vec{X}_2, \dots, \vec{X}_m\}$, where each $\vec{X}_i$ is an $N$-dimensional vector.

Consider a one-dimensional dynamical system for a single variable $x$:
\begin{equation}
    \frac{dx}{dt} = a x, \qquad x(0) = x_0,
\end{equation}
with $a \in \mathbb{R}$. Its analytical solution is:
\begin{equation}
    x(t) = e^{a t} x_0.
\end{equation}

Now, define an $N \times N$ projection matrix $\boldsymbol{\Omega}$ satisfying $\boldsymbol{\Omega}^2 = \boldsymbol{\Omega}$, and $\boldsymbol{\Omega}(\boldsymbol{I} - \boldsymbol{\Omega}) = 0$. We embed the system into a higher-dimensional space and study the dynamics of a single $N$-dimensional vector $\vec{X}_i$:
\begin{equation}
    \frac{d\vec{X}_i}{dt} = a \boldsymbol{\Omega} \vec{X}_i - \alpha (\boldsymbol{I} - \boldsymbol{\Omega}) \vec{X}_i, \qquad \vec{X}_i(0) = x_0 \vec{b},
\end{equation}
where $\alpha > 0$ and $\vec{b}$ is any vector satisfying $\boldsymbol{\Omega} \vec{b} \neq \vec{0}$.

Since the system is linear, its solution is:
\begin{equation}
    \vec{X}_i(t) = e^{[a \boldsymbol{\Omega} - \alpha (\boldsymbol{I} - \boldsymbol{\Omega})] t} \vec{X}_i(0) \approx e^{a \boldsymbol{\Omega} t} \vec{X}_i(0),
    \label{eq:1dex}
\end{equation}
for $t \gg 1/\alpha$. For any projector, the identity
\begin{equation}
    e^{a \boldsymbol{\Omega}} = \boldsymbol{I} + (e^a - 1) \boldsymbol{\Omega}
    \label{eq:projident}
\end{equation}
holds, yielding the asymptotic solution:
\begin{equation}
    \vec{X}_i(t) \approx (\boldsymbol{I} - \boldsymbol{\Omega}) \vec{X}_i(0) + x(t) \boldsymbol{\Omega} \vec{b}.
    \label{eq:2dex}
\end{equation}
Projecting this expression:
\begin{equation}
    \boldsymbol{\Omega} \vec{X}_i(t) \approx x(t) \boldsymbol{\Omega} \vec{b},
    \label{eq:1dsol}
\end{equation}
reveals that the original scalar dynamics is replicated across the projective subspace.

We can also recover the original dynamics via averaging. Taking the average over $\vec{X}_i(t)$ using the one-vector $\vec{1}$:
\begin{equation}
    \frac{1}{N} \vec{1}^T \boldsymbol{\Omega} \vec{X}_i(t) \approx x(t) \frac{1}{N} \vec{1}^T \boldsymbol{\Omega} \vec{b}.
\end{equation}
Choosing $\vec{b}$ such that:
\begin{equation}
    \frac{1}{N} \vec{1}^T \boldsymbol{\Omega} \vec{b} = \frac{1}{N} \sum_{\alpha,\beta=1}^N \Omega_{\alpha\beta} b_\beta = 1,
\end{equation}
ensures that:
\begin{equation}
    x(t) = \frac{1}{N} \vec{1}^T \boldsymbol{\Omega} \vec{X}_i(t) = \frac{1}{N} \sum_{\alpha,\beta=1}^N \Omega_{\alpha\beta} X_{i,\beta}(t).
\end{equation}

Here we use Greek symbols when utilizing the PrEDS method to indicate individual elements of the lifted space.  
We now generalize to nonlinear systems. Given a system $\frac{d\vec{x}}{dt} = \vec{f}(\vec{x})$, or equivalently $\vec{f}(\vec{x}) = -\nabla_{\vec{x}} V(\vec{x})$ if a potential $V$ is defined, the embedded dynamics under PrEDS reads:
\begin{equation}
    \frac{d\vec{X}_i}{dt} = \mathbf{\Omega} \vec{F}_i(\vec{X}_1, \dots, \vec{X}_m) - \alpha(\mathbf{I} - \mathbf{\Omega}) \vec{X}_i,
    \label{eq:PrEDS_dynamics2}
\end{equation}
where $\mathbf{\Omega}$ is a normalized mean-field projector, $\Omega_{\alpha\beta} = \frac{1}{N}$, and $\vec{F}_i$ is the lifted version of $f_i(\vec{x})$, applied row-wise to the $N \times m$ matrix of $\{\vec{X}_1, \vec{X}_2, \dots, \vec{X}_m\}$. The off-plane decay term ensures convergence to the projective subspace.

This is a particular case of a more general form. However in the case of the mean field projector, $\vec{F}_i$ corresponds to a commutative map as defined in \cite{CARAVELLI2023133747}, where the projection acts on the output of the function rather than its arguments. The Banality Lemma derived in \cite{CARAVELLI2023133747} guarantees that fixed points of the original system are preserved under the projection. It follows that the low-dimensional dynamics can be retrieved via averaging:
\begin{equation}
    \langle \vec{X}_i \rangle = \frac{1}{N} \sum_\alpha \vec{X}_{i,\alpha},
\end{equation}
where 
In the remainder of this work, we will apply PrEDS using network projectors $\boldsymbol{\Omega}_A$, $\boldsymbol{\Omega}_B$, and the mean-field projector $\mathbf{\Omega}$ via the commutative map formalism.

\section{Physical and Biological Systems as PrEDS}
\label{appendix:BioPhysSystems}

\subsection{Circuit-Based Dynamics and Memristive Networks}
\label{appendix:memristors}

This appendix provides the full derivations underlying the projector-based formulation of resistive and memristive circuits discussed in the main text, and clarifies how these structures support a PrEDS description. We begin with current–voltage relations in resistive networks, then derive the projected memristor dynamics, including alternative parametrizations and inversion identities needed for efficient simulation and analysis.

\subsection*{Kirchhoff Constraints and Projector Dynamics}

Electrical circuits satisfy Kirchhoff’s Current Law (KCL) and Voltage Law (KVL), which correspond to the nullspaces of the incidence and cycle matrices $B$ and $A$, respectively. These constraints allow one to construct oblique projection operators that encode node- and loop-based constraints.

For a resistive network with external sources, the current and voltage relations are:
\begin{subequations}
\begin{align}
    \vec{i} &= G \vec{v} + \vec{j}_\text{ext}, \\
    \vec{v} &= R \vec{i} + \vec{s},
\end{align}
\end{subequations}
where $G = R^{-1}$ is the conductance matrix. Using KVL and the projector $\Omega_{A/R} = A^T (A R A^T)^{-1} A$, the power dissipation becomes:
\begin{align}
2\mathcal{D} &= \vec{i}^T R A^T (A R A^T)^{-1} A \vec{v} \nonumber \\
&= (\vec{v} - \vec{s})^T A^T (A R A^T)^{-1} A \vec{v} \nonumber \\
&= (A \vec{v})^T (A R A^T)^{-1} (A \vec{v}) - \vec{s}^T \Omega_{A/R} \vec{v} \nonumber \\
&= -\vec{s}^T A^T (A R A^T)^{-1} A \vec{v}, \\
\Rightarrow \quad \vec{i} &= -A^T (A R A^T)^{-1} A \vec{s}.
\end{align}

Similarly, applying KCL with $\Omega_{B/G} = B^T (B G B^T)^{-1} B$ gives:
\begin{align}
2\mathcal{D} &= \vec{i}^T B^T (B G B^T)^{-1} B G \vec{v} \nonumber \\
&= \vec{i}^T B^T (B G B^T)^{-1} B  (\vec{i} - \vec{j}_\text{ext}) \nonumber \\
&= - \vec{i}^T B^T (B G B^T)^{-1} B  \vec{j}_\text{ext}, \\
\Rightarrow \quad \vec{v} &= -B^T (B G B^T)^{-1} B \vec{j}_\text{ext}.
\end{align}

These expressions form the foundation for the dynamical generalizations that follow.

\subsection*{Memristor Network Dynamics}

Memristors are resistive devices whose internal resistance $R(x)$ depends on a memory variable $x(t) \in [0, 1]$, which corresponds to high ($R_\text{off}$) and low ($R_\text{on}$) resistance states.. For the linear metal-oxide model, one writes:
\begin{align}
    R(x) &= R_{\text{on}} x + R_{\text{off}} (1 - x), \\
    \frac{dx}{dt} &= \frac{R_{\text{off}}}{\beta} i(t) - \alpha x(t),
    \label{eq:memr1}
\end{align}
with $i(t)$ the local current,  $\alpha$ is a decay rate (units of inverse time), and $\beta$ is an inverse learning rate (units of voltage-time). We define the scaling factor $\xi = \frac{R_\text{off} - R_\text{on}}{R_\text{on}}$. The resistance can also be written in scaled form as:
\begin{equation}
    R(x) = R_\text{off} (1 - \chi x), \qquad \chi = \frac{R_\text{off} - R_\text{on}}{R_\text{off}}.
\end{equation}

Let $X = \text{diag}(\vec{x})$ represent the network state. Using the projection operator $\Omega_A$, the current vector becomes:
\begin{equation}
    \vec{i} = -R_\text{off}^{-1} \left(I - \chi \Omega_A X \right)^{-1} \Omega_A \vec{s}.
    \label{eq:isol2}
\end{equation}
Combining this with the update law for $x(t)$ yields the projected dynamics:
\begin{equation}
    \frac{d\vec{x}}{dt} = -\frac{1}{\beta} \left(I - \chi \Omega_A X \right)^{-1} \Omega_A \vec{s} - \alpha \vec{x}.
\end{equation}
Here we have incorporated the constant $\Roff^{-1}$ into the learning rate $\beta^{-1}$. 
Alternatively, if voltage sources $\vec{S}$ are defined such that the memristor polarity aligns with $\vec{S}$, then:
\begin{equation}
    \frac{d\vec{x}}{dt} = \frac{1}{\beta} \left(I - \chi \Omega_A X \right)^{-1} \Omega_A \vec{S} - \alpha \vec{x},
    \label{eqn:BiasControlled1}
\end{equation}
or, in projected PrEDS form:
\begin{equation}
    \frac{d\vec{x}}{dt} = \Omega_A \left( \frac{1}{\beta} \left(I - \chi \Omega_A X \right)^{-1} \Omega_A \vec{S} - \alpha \vec{x} \right) - \alpha (I - \Omega_A) \vec{x}.
\end{equation}
The dynamics are governed by the relaxation to the minima. The minima is determined by the embedding via the projection operator. It is not obvious when the dynamics of different embeddings coincide, such that the minima determined by different embedding processes are the same.

\subsection*{Mean-Field Embedding of Memristor Dynamics}

Lifting $\vec{x}$ to a set of vectors $\{ \vec{X}_i \}$ using the PrEDS formalism, and applying the mean-field projector $\mathbf{\Omega}$, we write:
\begin{align}
 \frac{d\vec{X}_i}{dt} = \frac{1}{\beta} \mathbf{\Omega} \left[(I - \chi \Omega_A \, \text{Diag}(\{\vec{X}\}))^{-1} \Omega_A \vec{S}\right]_i - \alpha \mathbf{\Omega} \vec{X}_i - \alpha^\star (\mathbf{I} - \mathbf{\Omega}) \vec{X}_i.
 \label{eqn:PrEDS_memristor}
\end{align}
Here, $\vec{X}_i$ is a column vector in the higher-dimensional space. Each of the $N$ replicas systems in $\{X\}$ is drawn from the same random distribution as our original system $\vec{x}$, $m$ values drawn from a uniform distribution from $[0,1)$. This will be the case in each networks systems discussed.

\subsection*{Alternate Parametrizations}

An alternative “flipped” parametrization is given by:
\begin{align}
    R(x) &= \Ron(1 - x) + \Roff x, \\
    \frac{dx}{dt} &= \alpha x - \frac{\Ron}{\beta} i(t).
    \label{eq:memr2}
\end{align}
 Though algebraically similar, this model inverts the location of $R_{\text{on}}$ and $R_{\text{off}}$. The two parametrizations are dynamically related via:
\begin{equation}
\alpha \leftrightarrow -\alpha, \quad \beta \leftrightarrow -\beta, \quad \xi \leftrightarrow -\chi,
\end{equation}
but they are not physically equivalent.

\subsection*{Flipped Model: Mean-Field Embedding} 

In the flipped model, the dynamics under mean-field embedding become:
\begin{equation}
\frac{d\vec{x}}{dt} = -\frac{1}{\beta} (I + \xi \Omega_A X)^{-1} \Omega_A \vec{S} + \alpha \vec{x}.
\end{equation}
Lifted via PrEDS with $\mathbf{\Omega}$:
\begin{equation}
\frac{d\vec{X}_i}{dt} = -\frac{1}{\beta} \mathbf{\Omega} (I + \xi \Omega_A \text{Diag}(\vec{X}_i))^{-1} \Omega_A \vec{S} + \alpha \mathbf{\Omega} \vec{X}_i - (\mathbf{I} - \mathbf{\Omega}) \vec{X}_i.
\end{equation}
This matches the general PrEDS form and demonstrates how the projector governs learning trajectories even under alternative parameterizations.

\subsection*{Series Voltage Derivation and Inversion Identity}

Using the Woodbury matrix identity,
\begin{equation}
    (P+Q)^{-1} = \sum_{k=0}^\infty (-P^{-1} Q)^k P^{-1},
    \label{eqn:expansion}
\end{equation}
with $P = A A^T$, $Q = A Z A^T$, and $Z = \text{diag}(\vec{x})$, one arrives at:
\begin{align}
A^T (A A^T + A Z A^T)^{-1} A &= \sum_{k=0}^\infty (-1)^k (\Omega_A Z)^k \Omega_A \nonumber \\
&= \Omega_A (I + Z)^{-1} \Omega_A = (I + \Omega_A Z)^{-1} \Omega_A.
\label{eqn:InvDerivation}
\end{align}

\subsection{Flow Networks}

\label{appendix:flownets}

We now derive the PrEDS representation for biological flow networks, using the example of \textit{Physarum polycephalum}, a slime mold whose adaptive behavior is governed by local feedback based on mass flow. The dynamic remodeling of its tube-like structure can be modeled as a flow network governed by pressure gradients and conservation constraints.

\subsection*{Modeling Flow and Pressure}

The relationship between pressure and flow is captured by Poiseuille’s law:
\begin{equation}
Q_{ij} = \frac{\pi \eta}{128} \frac{D_{ij}^4}{L_{ij}} (p_i - p_j),
\label{eqn:poiseuille_flow}
\end{equation}
where $D_{ij}$ and $L_{ij}$ are the diameter and length of a tube, $p_i$, $p_j$ the pressures at connected nodes, and $\eta$ is the fluid viscosity. The conductance $g_{ij} = \frac{\pi \eta}{128} \frac{D^4_{ij}}{L_{ij}}$ allows the flow to be written analogously to Ohm’s law:
\begin{equation}
Q_{ij} = g_{ij} (p_i - p_j).
\end{equation}

Flow conservation at each node is expressed as:
\begin{equation}
\sum_i Q_{ij} = P_0 (\delta_{bj} - \delta_{Sj}),
\label{eqn:delta_flow}
\end{equation}
with $P_0$ the total input/output flow, and $b$, $S$ denoting body and sink nodes. This structure parallels KCL, with effective current conservation at each junction.

To account for pressure sources and sinks, we embed the graph $G$ into an extended graph $G' = G \cup G_e$, introducing a virtual ground node. This allows all potentials to be treated relative to a reference node. 
The pressure drop across each edge $\beta$ in $G'$ becomes:
\begin{equation}
\Delta V_{\beta} =
\begin{cases}
\frac{Q_{\beta}}{g_{\beta}} & \beta \in G \\
p_i - p_0 & \beta \in G_e
\end{cases}
\end{equation}
with conductance $g_\beta$ defined by tube geometry.

\subsection*{Network Resistance and Adaptive Remodeling}

Following \cite{Tero_Science_2010}, we model adaptation by associating a memory variable $x_{ij} \in [0,1]$ to each edge. This variable governs the effective resistance of the tube:
\begin{align}
R_{ij}(x) = \begin{cases}
\frac{1}{\eta'} \frac{1}{d_0^4} (L_{\text{min}} x + L_{\text{max}} (1 - x)) & \text{(length varies)} \\
\frac{1}{\eta'} l_0 (D_{\text{max}}^{-4} x + D_{\text{min}}^{-4} (1 - x)) & \text{(diameter varies)}
\end{cases}.
\label{eqn:slimeresistance}
\end{align}
In these networks,  $x = 0$ corresponds to maximum resistance (e.g., elongated or narrow tubes), and $x = 1$ to minimum resistance (e.g., short or wide tubes).

Adaptation dynamics are modeled by:
\begin{equation}
\frac{dx_{ij}}{dt} = \frac{1}{\beta} |Q_{ij}| - \kappa x_{ij},
\label{eqn:flow_independent}
\end{equation}
where $\kappa$ is a decay parameter, and $\beta$ a scaling factor.

\subsection*{Projected Flow Dynamics}

Using circuit theory, the network flow vector $\vec{Q}$ can be expressed in terms of source potentials via:
\begin{align}
\vec{Q} &= -A^T (A R A^T)^{-1} A \, \Delta \vec{V}_\text{source} \\
&= -\frac{1}{R_{\text{off}}} (I - \chi \boldsymbol{\Omega}_A X)^{-1} \boldsymbol{\Omega}_A \Delta \vec{V}_\text{source},
\label{eqn:flow_Lchanging}
\end{align}
where $X = \text{diag}(\vec{x})$ and $\chi$ parameterizes the resistance range.

The nonlinear update rule for $x$ involves the absolute value of $\vec{Q}$ and cannot be expanded directly. Still, we can write the PrEDS-compatible form for individual and mean-field lifted dynamics:
\begin{align}
\frac{d\vec{x}}{dt} &= -\kappa \vec{x} + 
\begin{cases}
-\frac{1}{\beta\Roff} \left((I - \chi \boldsymbol{\Omega}_A X)^{-1} \boldsymbol{\Omega}_A \Delta \vec{V}_\text{source}\right)_{i} ,& Q_i > 0 \\
+\frac{1}{\beta\Roff} \left((I - \chi \boldsymbol{\Omega}_A X)^{-1} \boldsymbol{\Omega}_A \Delta \vec{V}_\text{source}\right)_{i} ,& Q_i \leq 0
\end{cases}, \\
\frac{d\vec{X}_i}{dt} &= -\kappa \mathbf{\Omega} \vec{X}_i - \alpha^\star (\mathbf{I} - \mathbf{\Omega}) \vec{X}_i + \frac{1}{\beta\Roff}
\begin{cases}
\mathbf{\Omega} \left[ -\cdots \right]_{i\beta} ,& Q_\beta > 0 \\
\mathbf{\Omega} \left[ +\cdots \right]_{i\beta} ,& Q_\beta \leq 0
\end{cases}
\end{align}

A more compact mean-field expression is:
\begin{equation}
\frac{d\vec{X}_i}{dt} = -\kappa \mathbf{\Omega} \vec{X}_i + \frac{1}{\beta\Roff} \mathbf{\Omega} \left[\,\left| (I - \chi \boldsymbol{\Omega}_A \text{Diag}(\{\vec{X}\}))^{-1} \boldsymbol{\Omega}_A \Delta \vec{V}_\text{source} \right|\,\right]_i - \alpha^\star (\mathbf{I} - \mathbf{\Omega}) \vec{X}_i.
\end{equation}

\begin{figure}[ht]
\centering
\includegraphics[width=.9\textwidth]{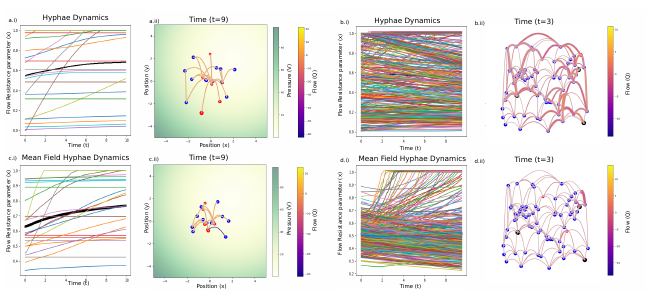}
\caption{A network of dynamical fluid flow, modeled as a network of tubes, with unit-less parameters $(\kappa,\beta,\chi,\alpha^\star)=(0.001,0.001,0.9,1)$ for (a) and (c), and $(0.05,0.015,0.9,1,1)$ for (b) and (d). (a,c) show edge length adaptation; (b,d) show diameter adaptation. Pressure field gradients are shown in green; nodes are color-coded by role.}
\label{fig:Slime}
\end{figure}

\subsection*{Flow Network Simulation Results}
Figure \ref{fig:Slime} shows the evolution of flow networks under different adaptation schemes, wherein the length ((a) and (c)) or diameter ((b) and (d)) adapt due to flow.  Evolution of the edge length  models a network exploring a landscape. Simulations involve optimizing the network configuration as the edge lengths evolve. The local position of source nodes (red nodes) determines the pressure at the nodes, a potential energy gradient is shown in green in the network configuration, (a.ii), (c.ii), analogous to a gradient of chemicals. The network readjusts to the minima of the gradient as the network is dissipative. An active network such as \textit{Physarum polycephalum} can explore a chemical gradient, e.g., to search for food by increasing the length of network edges. This is modeled by changing the sign of eq.~\eqref{eqn:slimeresistance}.

Evolution of the edge diameter models a fixed flow network, (b) and mean field dynamics (d) both show non-monotonicity in network resistance under driving. In contrast to the memristive networks in Fig.~\ref{fig:Memcircuit}, resistance in flow networks obtains both high and low resistance values. Network configurations (b.ii) and (d.ii) demonstrate large diameter tubes near fluid sinks (black nodes). In the supplementary material we show more snapshots of the network evolution under driving for both dynamical systems.
Examining the mean field model (c) and (d) it appears the mean field model adjusts parameters that the original dynamics do not evolve, for example comparing (a) and (c), the length of many edges in the original dynamics do not evolve under forcing, and thus at comparable time-points the network in (a.ii) is more spread out than in (c.ii). Comparing (b) and (d) under the mean field dynamics more resistance values are evolving at the same time, where in the original dynamics, the resistance in many edges is not changing significantly. The mean field projector is able to sample a wider range of configuration space to drive the system to the minima.

\begin{figure}[ht]
\centering
\includegraphics[width=.8\textwidth]{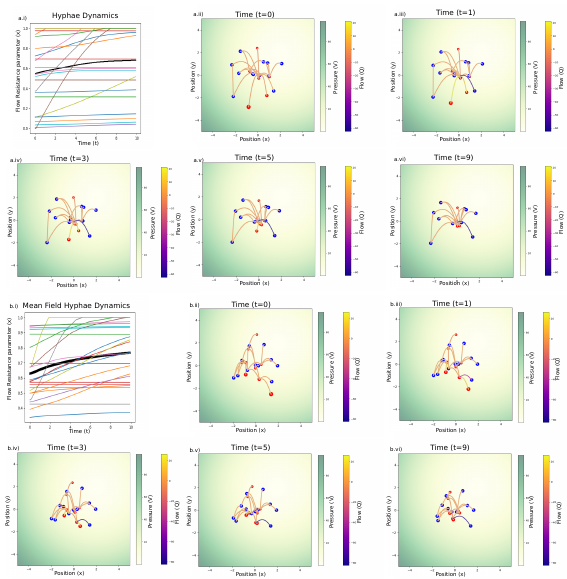}
\caption{(a) and (b) show independent and mean-field networks with edge length adaptation. (a.i, b.i) Resistance evolution over time. Mean-field dynamics produce faster convergence and fewer stagnant edges.}
\label{fig:Flow_Lengths}
\end{figure}

\begin{figure}[ht]
\centering
\includegraphics[width=.8\textwidth]{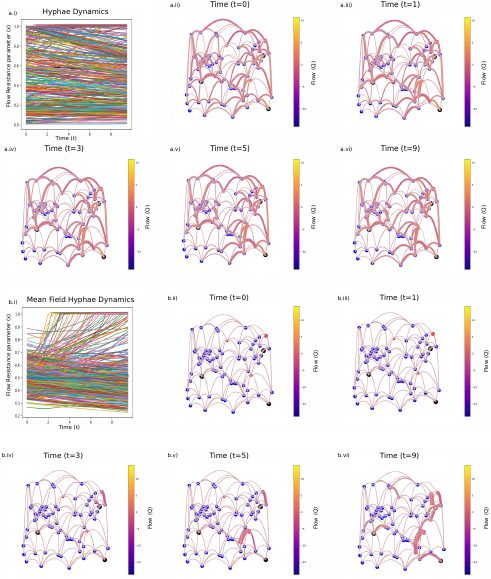}
\caption{(a) and (b) Independent and mean field dynamics under diameter adaptation. (a.i, b.i) show time evolution of edge resistances.}
\label{fig:Flow_Diameter}
\end{figure}

The adaptation dynamics under both individual and mean-field embeddings are shown in Figures~\ref{fig:Flow_Lengths} and \ref{fig:Flow_Diameter}. In the case of dynamic edge length (Fig.~\ref{fig:Flow_Lengths}), node positions adjust to accommodate changing resistances. In contrast, diameter adaptation (Fig.~\ref{fig:Flow_Diameter}) retains node locations, with flow magnitudes and resistance values evolving over time.

\subsection{Hookean Networks and Projection-Based Adaptation}
\label{appendix:springnetworks}

This section presents the full derivation of the PrEDS formulation for mechanical networks of Hookean springs. These systems exhibit local conservation of forces and displacements, and their dynamics can be naturally analyzed using network projectors. We develop here the operator structure, energy balances, and memory dynamics associated with such elastic networks.

We consider networks composed of elastic elements, idealized as Hookean springs, connected at nodes. For clarity, we restrict to planar networks, although the formalism generalizes to three-dimensional systems.

Following the PrEDS procedure, we first identify the conserved quantities: the forces and displacements must satisfy local conservation laws. At each node, the total force vanishes:
\begin{subequations}
\begin{align}
\sum_{ij} \vec{F}_{ij} &= 0, \\
\sum_{ij} -k_{ij}(\vec{x}_{ij} - \vec{X}^\text{eq}_{ij}) &= 0,
\end{align}
\end{subequations}
where $k_{ij}$ is the spring constant, $\vec{x}_{ij}$ the current displacement vector between nodes $i$ and $j$, and $\vec{X}^\text{eq}_{ij}$ the corresponding equilibrium vector.

Moreover, around any closed cycle $c$ in the network, the net displacement must vanish:
\begin{subequations}
\begin{align}
\sum_{e_{ij} \in c} \vec{x}_{ij} &= 0, \\
\sum_{e_{ij} \in c} \frac{d\vec{x}_{ij}}{dt} &= 0.
\label{eqn:SpringLoop}
\end{align}
\end{subequations}
Equation \eqref{eqn:SpringLoop} reflects that while edge vectors can evolve in time, the topological connectivity of the graph remains fixed. These constraints are Galilean invariant; a uniform velocity shift across all nodes does not affect the internal dynamics.

To enforce conservation laws, we define the incidence and cycle matrices $B$ and $A$. Force and displacement components in the $x$ and $y$ directions satisfy:
\begin{equation}
B \vec{F}^x = 0, \quad B \vec{F}^y = 0, \quad A \vec{x}^x = 0, \quad A \vec{x}^y = 0.
\end{equation}

Power exerted by the spring network is given by the inner product of conserved quantities:
\begin{subequations}
\begin{align}
P &= \sum_{e_{ij} \in \mathcal{G}} -\vec{v}_{ij}^T \cdot \vec{F}_{ij}, \\
&= \sum_{e_{ij} \in \mathcal{G}} \frac{d\vec{x}_{ij}^T}{dt} \cdot (\vec{x}_{ij} - \vec{X}^\text{eq}_{ij}) k_{ij}.
\end{align}
\end{subequations}

For planar spring networks, we analyze power and energy in each component:
\begin{align}
P^\mu &= \sum_{e_{ij} \in \mathcal{G}} \frac{d\vec{x}_{ij}^{\mu\ T}}{dt} \, k_{ij} (\vec{x}^\mu_{ij} - \vec{X}^{\mu \text{eq}}_{ij}), \\
E^\mu &= \sum_{e_{ij} \in \mathcal{G}} \frac{1}{2} \Delta \vec{x}^\mu_{ij}{}^T \, k_{ij} \Delta \vec{x}^\mu_{ij},
\end{align}
where $\mu \in \{x, y\}$ and $\Delta \vec{x}_{ij} = \vec{x}_{ij} - \vec{X}^\text{eq}_{ij}$ is the displacement from equilibrium.

We write power dissipation in component form using:
\begin{equation}
2\mathcal{D}^\mu = -\frac{1}{2} \frac{d}{dt} \left( \Delta \vec{x}^{\mu\, T} K \Delta \vec{x}^\mu \right) = -\frac{1}{2} \frac{d}{dt} \left( \Delta \vec{x}^{\mu\, T} \vec{F}^\mu \right),
\end{equation}
where $K$ is the diagonal matrix of spring constants. The displacement vector is given in terms of a biased force:
\begin{equation}
\Delta \vec{x}^\mu = B^T (B K B^T)^{-1} B \vec{F}^\mu_\text{bias}.
\end{equation}

We introduce a memory parameter $z \in [0,1]$ to encode adaptive spring constants:
\begin{equation}
k = k_\text{max} z + k_\text{min}(1 - z).
\end{equation}
Then, the displacements in $x$ and $y$ are given by:
\begin{subequations}
\begin{align}
\Delta \vec{x} &= -B^T (B K B^T)^{-1} B \vec{F}_\text{bias} \cos(\theta), \label{eqn:Spring_DeltaX} \\
\vec{x} &= -B^T (B K B^T)^{-1} B \vec{F}_\text{bias} \cos(\theta) - \vec{X}^\text{eq} \cos(\theta) \\
&= -\frac{1}{k_\text{max}} (I + \chi \Omega_B Z)^{-1} \Omega_B \vec{F}_\text{bias} \cos(\theta) - \vec{X}^\text{eq} \cos(\theta), \nonumber
\end{align}
\end{subequations}
\begin{subequations}
\begin{align}
\Delta \vec{y} &= -B^T (B K B^T)^{-1} B \vec{F}_\text{bias} \sin(\theta), \label{eqn:Spring_DeltaY} \\
\vec{y} &= -B^T (B K B^T)^{-1} B \vec{F}_\text{bias} \sin(\theta) - \vec{X}^\text{eq} \sin(\theta) \\
&= -\frac{1}{k_\text{max}} (I + \chi \Omega_B Z)^{-1} \Omega_B \vec{F}_\text{bias} \sin(\theta) - \vec{X}^\text{eq} \sin(\theta), \nonumber
\end{align}
\end{subequations}
where $\theta_{ij} = \arctan(y_{ij} / x_{ij})$ is the angle of the spring element $e_{ij}$, and $\vec{F}_\text{bias} \cos(\theta)$ and $\vec{F}_\text{bias} \sin(\theta)$ are vectors of force components along each edge direction.

\begin{figure}[ht]
\centering
\includegraphics[width=.9\textwidth]{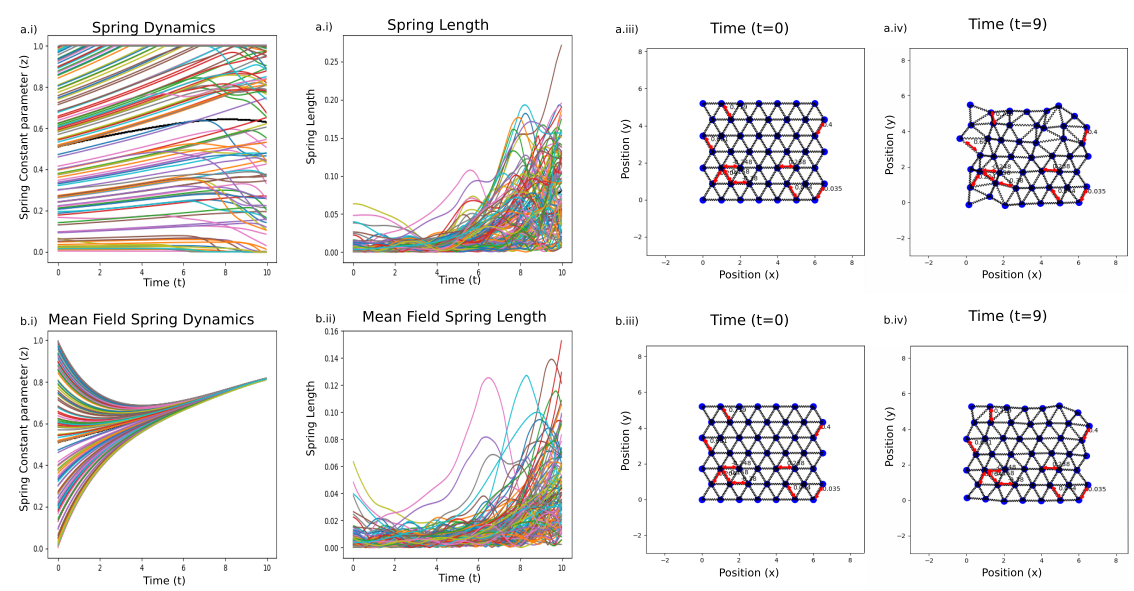}
\caption{ A network of coupled springs under a force aligned with the spring with unit-less parameters $(\alpha,\beta,\chi,\alpha^\star, ||X^\text{eq}||)=( 0.1, 10,0.95 ,1,1)$. (a.i) and (b.i) 
 Individual and mean field dynamics of the spring constant. (a.ii) and (b.ii) Length of the springs under biasing. Spring network configuration ordered and at equilibrium (a.i) and (b.i), and after biasing (a.iv) and (b.iv) are shown, bias forces as shown as red arrows along edges.}
\label{fig:Spring}
\end{figure}

This formulation shows that spring networks, like fluid or electrical systems, can be described using projection operators, with network topology encoded in $B$ and system memory in $Z$. The resulting dynamics satisfy local conservation laws and allow mean-field treatment via the PrEDS framework.

 We now identify the dynamical equations for the spring network. While mass is conserved, we allow the spring constants to evolve dynamically, following an energy-based update rule of the form $\frac{dk}{dt} \propto -\frac{\partial E}{\partial k}$. This modeling approach is well-established in the study of adaptive mechanical circuits \cite{patil2023selflearningmechanicalcircuits, Stern_PRX_2021}. The dynamics are given by:
\begin{subequations}
\begin{align}
    \frac{d\vec{z}}{dt} &= \alpha \vec{z} - \beta \Delta \vec{l}^2 \\
    &= \alpha \vec{z} - \beta \left( \Delta \vec{x}^2 + \Delta \vec{y}^2 \right) \\
    &= \alpha \vec{z} - \beta \left( \left(B^T (B K B^T)^{-1} B \Omega_B \vec{F}_\text{bias} \cos\theta \right)^2 
    + \left(B^T (B K B^T)^{-1} B \Omega_B \vec{F}_\text{bias} \sin\theta \right)^2 \right),
\end{align}
\end{subequations}
where $\Delta \vec{l}^2 = \Delta \vec{x}^2 + \Delta \vec{y}^2$ is the squared displacement.

Introducing the scaling $\chi = \frac{k_\text{max} - k_\text{min}}{k_\text{max}}$, the equation becomes:
\begin{subequations}
\begin{align}
    \frac{d\vec{z}}{dt} &= \alpha \vec{z} - \frac{\beta}{k_\text{max}} \left( 
    \left[(I + \chi \Omega_B Z)^{-1} \Omega_B \vec{F}_\text{bias} \cos\theta \right]^2 
    + \left[(I + \chi \Omega_B Z)^{-1} \Omega_B \vec{F}_\text{bias} \sin\theta \right]^2 \right).
\end{align}
\label{eqn:SpringDynamicalEqn}
\end{subequations}

This formulation expresses the spring dynamics directly in terms of orthogonal projection operators. Using the mean-field projector, we rewrite Eq.~\eqref{eqn:SpringDynamicalEqn} as:
\begin{subequations}
\begin{align}
    \frac{d\vec{Z}}{dt} &= \alpha \mathbf{\Omega} \vec{Z} \\
    &\quad - \frac{\beta}{k_\text{max}} \mathbf{\Omega} \left( 
    \left[(I + \chi \Omega_B \text{Diag}(\vec{Z}))^{-1} \Omega_B \vec{F}_\text{bias} \cos\theta \right]^2 
    + \left[(I + \chi \Omega_B \text{Diag}(\vec{Z}))^{-1} \Omega_B \vec{F}_\text{bias} \sin\theta \right]^2 \right) \nonumber \\
    &\quad - \alpha^\star (\mathbf{I} - \mathbf{\Omega}) \vec{Z}.
\end{align}
\label{eqn:PrEDS_SpringDynamicalEqn}
\end{subequations}

In our simulations, Eqs.~\eqref{eqn:Spring_DeltaX} and \eqref{eqn:Spring_DeltaY} are used to compute both the spring lengths $\Delta l$ and local angles $\theta$, which are then updated iteratively to simulate the time evolution of the spring constants.

Figure~\ref{fig:Spring} shows the evolution of spring constants, spring lengths, and spring configurations under applied bias. Panels (a.i) and (b.i) display the dynamics of individual and mean-field spring constants, respectively. While individual dynamics exhibit non-monotonic behavior due to rotation and local rearrangements, the mean-field dynamics are monotonic, steadily increasing the average spring constant. In panels (a.ii) and (b.ii), we observe that individual dynamics result in longer spring lengths compared to the mean-field case, which produces more compact, less buckled configurations due to uniformly increased stiffness.

\subsection*{Hookean Network Simulation Results}
In our simulations, Eqs.~\eqref{eqn:Spring_DeltaX} and \eqref{eqn:Spring_DeltaY} are used to compute both the spring lengths $\Delta l$ and local angles $\theta$, which are then updated iteratively to simulate the time evolution of the spring constants.

\begin{figure}[ht]
\centering
\includegraphics[width=.9\textwidth]{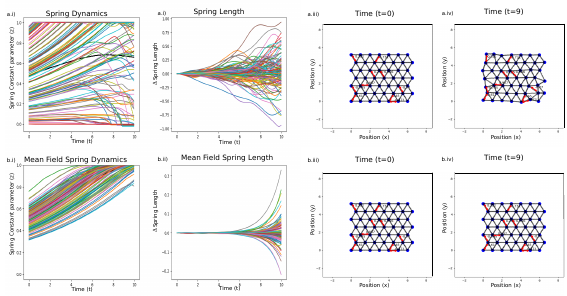}
\caption{Adaptive spring network under forcing. Panels (a) and (b) show individual vs. mean-field dynamics. (a.i–b.i): time evolution of spring stiffness; (a.ii–b.ii): spring length deviations; (a.iii–a.iv) and (b.iii–b.iv): configurations before and after adaptation.}
\label{fig:Spring}
\end{figure}
Figure~\ref{fig:Spring} shows the evolution of spring constants, spring lengths, and spring configurations under applied bias. Panels (a.i) and (b.i) display the dynamics of individual and mean-field spring constants, respectively. While individual dynamics exhibit non-monotonic behavior due to rotation and local rearrangements, the mean-field dynamics are monotonic, steadily increasing the average spring constant. In panels (a.ii) and (b.ii), we observe that individual dynamics result in longer spring lengths compared to the mean-field case, which produces more compact, less buckled configurations due to uniformly increased stiffness. In the spring networks local reconfigurations result from local variation in the spring constants, by averaging over multiple configurations are stable spring configuration persists for longer.

\subsection*{Power Dissipation via Projectors}

We now recover the projector structure from energy dissipation. Starting from:
\begin{align}
\partial_t(\Delta \vec{x}^{\mu\,T} K \Delta \vec{x}^\mu)
&= \partial_t(\Delta \vec{x}^{\mu\,T} K B^T (BKB^T)^{-1} B K \Delta \vec{x}^\mu), \\
&= -2 \frac{d \vec{x}^{\mu\,T}}{dt} K B^T (B K B^T)^{-1} B \vec{F}_{\text{bias}}^\mu.
\end{align}
This establishes:
\begin{align}
\Delta \vec{x}^{\mu} = -B^T (B K B^T)^{-1} B \vec{F}^\mu_\text{bias}.
\end{align}

We also compute:
\begin{align}
\partial_t(\vec{F}^{\mu\,T} \Delta \vec{x}^\mu)
= \partial_t\left( \vec{F}^{\mu\,T} K^{-1} A^T (A K^{-1} A^T)^{-1} A K^{-1} \vec{F}^\mu \right),
\end{align}
implying:
\begin{align}
\vec{F}^{\mu\,T} \Delta \vec{x}^\mu = \vec{F}^{\mu\,T} K^{-1} A^T (A K^{-1} A^T)^{-1} A K^{-1} \vec{F}^\mu + C.
\end{align}

We set the constant $C=0$. This recovers the duality between edge-based and loop-based energy flows, again showing that projectors encode the structure of reversible and dissipative contributions.

\section{Particle swarms}
\label{appendix:Swarm}

\subsection{From PrEDS to particle dynamics}

We first explicitly write out the components of each vector in  eq.~\eqref{eq:PrEDS_dynamics}, which reads
\begin{equation}
    \frac{d \vec{X}_{i,\beta}}{dt} = \sum_{\theta=1}^{N} \frac{1}{N} f_i(X_{1,\theta},X_{2,\theta} ... X_{m,\theta}) - \alpha \sum_{\theta=1}^{N} (\delta_{\beta \theta} - \frac{1}{N}) X_{i,\theta}
\end{equation}
where $X_{i,\beta}$ is the $\beta$-th component of vector $\vec{X}_{i}$, we substitute the definition of particle position $\vec{r}_{\beta} = [X_{1,\beta},X_{2,\beta},...X_{m,\beta}]^T $ and get
\begin{align}
    \frac{d \vec{r}_{\beta,i}}{dt} &= \sum_{\theta=1}^{N} \frac{1}{N} f_i(\vec{r}_\beta) - \alpha\left(r_{\beta,i} -\frac{1}{N} \sum_\theta r_{\theta,i}\right) \\
    &= \frac{1}{N} \sum_{\theta=1}^{N}  f_i(\vec{r}_\beta) - \frac{\alpha}{N} \sum_\theta (r_{\beta,i}  - r_{\theta,i})
\end{align}
where $r_{\beta,i}$ is the i-th component of the position for particle number $\beta$, $\vec{r}_{\beta}$. By using $\vec{f}(\vec{r}_\beta) = -\nabla_{\vec{r}_\beta} V(\vec{r}_\beta)$ and $\vec{r}_\beta - \vec{r}_\theta = - \nabla_{\vec{r}_\beta} (|\vec{r}_\beta - \vec{r}_\theta|^2)$, we obtain eq.~\eqref{eq:PrEDS_particle_dynamics} in the main text.
\subsection{Coarse-grained continuum equation}
We first use the chain rule to calculate the time derivative,
\begin{align}
    \frac{d}{dt}f(\Vec{r}_\beta(t)) 
    &= \nabla_{\vec{r}_\beta} f(\Vec{r}_\beta(t))  \cdot \frac{d \Vec{r}_\beta(t) }{dt} 
    = \int d\vec{r} \ \delta(\vec{r} - \vec{r}_\beta) \nabla_{\vec{r}}f(\vec{r}) \cdot \frac{d \Vec{r}_\beta(t) }{dt}\\
    &= -\int d \vec{r} \ \rho_\beta (\Vec{r},t) \nabla_{\vec{r}} f(\vec{r}) \cdot \left(\frac{1}{N} \sum_{\theta}  \nabla_{\theta} V(\vec{r}_\theta) + \frac{\alpha}{N}\sum_{\theta} \nabla_{\vec{r}} U_H(\vec{r} - \vec{r}_{\theta})\right).
\end{align}
We then perform integration by parts to finally get eq.~\eqref{eq:dt_f}. 
After summing over the particle number, $\beta$, in eq.~\eqref{eq:dt_f}, we obtain the form
\begin{align}
    \partial_t \rho(\Vec{r},t)
    &=\nabla_{\vec{r}}  \cdot \left[\rho (\Vec{r},t)\left(\frac{1}{N} \sum_{\theta}  \nabla_{\vec{r}_\theta} V(\vec{r}_\theta) + \frac{\alpha}{N}\sum_{\theta} \nabla_{\vec{r}} U_H(\vec{r} - \vec{r}_{\theta})\right)\right] .
\end{align}
Using the properties of delta functions, we rewrite this equation as
\begin{align}
    \partial_t \rho(\Vec{r},t)
    &= \nabla_{\vec{r}}  \cdot \left[ \rho (\Vec{r},t)\left( \frac{1}{N} \sum_{\theta}  \int d\vec{y}\  \nabla_{\Vec{y}}  V(\vec{y}) \delta(\vec{y} - \vec{r}_\theta) + \frac{\alpha}{N}\sum_{\theta} \int d\vec{y} \ \nabla_{\vec{r}} U_H(\vec{r} - \vec{y}) \delta (\vec{y} - \vec{r}_{\theta})\right) \right]\\
    &= \nabla_{\vec{r}}  \cdot \left[ \rho (\Vec{r},t)\left( \frac{1}{N}  \int d\vec{y}\  \nabla_{\Vec{y}}  V(\vec{y}) \rho(\vec{y},t) + \frac{\alpha}{N} \nabla_{\vec{r}} \int d\vec{y} \  U_H(\vec{r} - \vec{y}) \rho (\vec{y},t)\right) \right]
\end{align}


\end{document}